%% file: main.tex
\documentclass[journal]{IEEEtran}
\IEEEoverridecommandlockouts
\usepackage{cite}
\usepackage{amsmath,amssymb,amsfonts}
\usepackage{algorithmic}
\usepackage{graphicx}
\usepackage{textcomp}
\usepackage{xcolor}
\usepackage{subcaption}
\usepackage{multirow}
\usepackage{rotating}
\usepackage{tabularx}
\usepackage{booktabs}
\usepackage{afterpage}
\usepackage{tablefootnote}
\usepackage[hidelinks]{hyperref}
\usepackage[dvipsnames]{xcolor}
\def\BibTeX{{\rm B\kern-.05em{\sc i\kern-.025em b}\kern-.08em
    T\kern-.1667em\lower.7ex\hbox{E}\kern-.125emX}}

\begin{document}

\title{Learning-Based vs Human-Derived Congestion Control: An In-Depth Experimental Study}

\author{\IEEEauthorblockN{Mihai Mazilu, Luca Giacomoni and George Parisis}
\\
\IEEEauthorblockA{\textit{School of Engineering and Informatics, University of Sussex} 
\\
\{m.mazilu, l.giacomoni, g.parisis\}@sussex.ac.uk}

}

\maketitle

\input{sections/Abstract}

\input{sections/Introduction}
\input{sections/Challenges}
\input{sections/Methodology}
\input{sections/Results}
\input{sections/Related_Work}
\input{sections/Discussion}

\bibliographystyle{IEEEtran}
\bibliography{bibliography-all-names}

\end{document}

%% file: sections/Abstract.tex
\begin{abstract}

Learning-based congestion control (CC), including Reinforcement-Learning (RL) approaches, promises efficient CC in a fast-changing networking landscape, where evolving communication technologies, applications and traffic workloads pose severe challenges to human-derived, fixed-at-deployment CC algorithms. Learning-based CC is in its early days and substantial research is required to understand existing limitations, identify research challenges and, eventually, yield deployable solutions for real-world networks. In this paper, we present a reproducible and systematic study of learning-based CC with the aim of highlighting strengths and uncovering fundamental limitations of the state-of-the-art, directly contrasting these approaches with widely deployed human-derived CC algorithms, namely TCP Cubic and BBR (version 3). We identify challenges in evaluating learning-based CC, establish a methodology for studying said approaches and perform large-scale experimentation focused on fairness and efficiency. We then show that existing RL based approaches can acquire available bandwidth while largely maintaining low latency. Finally, we highlight that all existing learning-based CC approaches underperform when the available bandwidth and end-to-end latency dynamically change, while remaining resistant to non-congestive loss. Our experimentation codebase and datasets are publicly available with the aim to galvanise the research community towards transparency and reproducibility, which have been recognised as crucial for researching and evaluating learning-based policies.

\end{abstract}

%% file: sections/Introduction.tex
\section{Introduction}
\label{introduction}


Reinforcement learning (RL) has gained substantial momentum in recent years, and increasingly sophisticated systems that span a wide array of applications, such as games \cite{mnih2013playing,vinyals2019grandmaster,silver2017mastering,silver2018general}, natural language processing \cite{ouyang2022training}, and even nuclear fusion \cite{degrave2022magnetic}, have been developed. Therefore, it has only been natural to see RL making its way into the networking realm, specifically in routing \cite{mammeri2019reinforcement}, video rate control \cite{huang2018qarc, mao2017neural}, network access \cite{wang2018deep, naparstek2018deep}, security \cite{nguyen2019deep, uprety2020reinforcement} and proactive caching \cite{zhu2018deep, he2017integrated}. Congestion control (CC) has recently been rethought through the lens of RL, promising efficient CC where fast-evolving communication technologies, applications, and traffic workloads pose substantial challenges to human-derived CC algorithms, such as Cubic \cite{cubic_paper} and BBR \cite{cardwell2016bbr}. Other learning-based CC approaches, such as Remy \cite{remy}, PCC Allegro \cite{pcc-allegro} and PCC Vivace \cite{pccvivace}, have also been explored.

It has been widely accepted that the experimental evaluation of CC algorithms is a challenging task due to the complexity of designing experiments that exercise CC under a wide range of network conditions and the breadth of CC performance metrics that must be taken into account in unison and not in isolation \cite{li2007experimental}. Learning-based CC approaches pose additional challenges because of the black-box nature of decision making (e.g., when deep RL is employed) or when the state space becomes very large to track decision making (e.g., in Remy). Consequently, machine-derived CC policies are difficult to reason about. The issues identified above are exacerbated by the diversity in the implementation of learning-based approaches and CC algorithms. For example, Remy \cite{remy}, TCP-Drinc \cite{xiao2019tcp} and SmartCC \cite{li2019smartcc} are implemented as simulation models. On the other hand, Aurora \cite{jay2019deep} is trained within a simulation environment and subsequently integrated into a user-space prototype. Orca \cite{abbasloo2020classic}, Astraea \cite{astraea} and Spine \cite{tian2022spine} perform RL model training and inference within a user space application, but the core of the CC is integrated within the Linux kernel. Sage \cite{sage} is trained on pre-collected traces from a diverse set of CC algorithms before being deployed within the Linux kernel. PCC Vivace comes with both a user- and kernel-space implementation, which behave very differently to each other.

We posit that existing literature on learning-based CC falls short when it comes to studying the behaviour of the CC policies and evaluating their performance; reproducibility of results has been underthought. Evaluation methodologies have been ad-hoc with experimentation conducted on emulated and `in the wild' environments. Network emulation can be very effective when experimenting with network protocols, but it is crucial that the complexities and limitations of the underlying queuing disciplines, buffering, hardware offloading and associated CPU overhead \cite{handigol2012reproducible} are carefully considered. In Aurora \cite{jay2019deep}, Orca \cite{abbasloo2020classic} and QTCP \cite{li2018qtcp} experimentation involves sending a single flow into a single path, and CC is only exercised when/if the sender's sending rate is higher than that capacity of the emulated path. In such scenarios, the congestion window (or sending rate) is quickly set to a value that the underlying RL agent considers (potentially locally) optimal and does not change much after that. Such experiments show how effective a CC algorithm is in capturing the available bandwidth in the absence of contending flows, but fall short when it comes to fairness and showing responsiveness in the face of hotspots. In \cite{abbasloo2020classic}, experimentation with different bandwidth-delay product (BDP) paths is un-systematic, with multiple BDP values tested when evaluating strengths in terms of eliminating bufferbloat, but only a single value when evaluating friendliness with TCP. `In the wild' environments (e.g., using GENI, and/or EC2 servers), are suitable for showcasing the feasibility of a CC approach and its compatibility with legacy CC, but pose profound limitations when it comes to interpreting and reproducing results \cite{yan2020learning}. For example, in \cite{abbasloo2020classic} a single flow is sent through inter- and intra-continental paths but it is unclear weather the flow experiences any congestion at all. When evaluating fairness, the characteristics of the randomly selected paths are not discussed, so it is impossible to reproduce the results. In \cite{astraea}, Astraea's convergence appears to be exceptional when evaluating within the parameters used to train the underlying model, however, in the code provided by the authors, we have observed that fair queuing has been configured in the emulated links; this skews the observed behaviour, which, as shown in this paper, is very good but not exceptional. Finally, it should be noted that there have not been any substantial comparative studies using simulation models and learning-based CC approaches. Frameworks like \cite{giacomoni2023raynet} and \cite{ns3-rl} open opportunities for developing an experimentation framework for RL-based CC using simulations.

In this paper, we expand on and update our initial study published in \cite{10621288} and present a reproducible and systematic study of learning-based CC with the aim of highlighting advances and uncovering fundamental limitations of the state-of-the-art. We identify challenges in evaluating learning-based CC, and devise a methodology and set of benchmark experiments to examine efficiency, fairness, and responsiveness in single- and multi-bottleneck topologies. We perform large-scale experimentation with publicly available RL-based CC models, namely \textit{Orca} \cite{abbasloo2020classic}, \textit{Sage} \cite{sage} and \textit{Astraea} \cite{astraea}, and \textit{PCC Vivace} \cite{pccvivace}, a learning rate-control algorithm. We compare those with \textit{Cubic} and \textit{BBR version 3}, both being human-derived and interpretable CC policies.

Three themes emerge from our expanded evaluation; (1) we have successfully reproduced Orca’s behaviour, matching earlier findings across different hardware configurations, with only limited variation; (2) we showcase that current state-of-the-art RL approaches struggle at \textit{capturing available bandwidth} in the presence of base RTT fluctuations compared to human-derived approaches; (3) generalisation breaks down; Astraea is fair only within its training parameters and fairness degrades sharply when operating outside those, while Sage becomes unstable outside its training range, oscillating its sending rate leading to slow convergence.

%% file: sections/Challenges.tex
\section{Challenges}
\label{challenges}

Previous work has identified several pitfalls in evaluating CC algorithms \cite{li2007experimental} such as conducting experiments that do not actually exercise congestion control, focusing on specific performance measures, and evaluating CC in a narrow range of network parameters. In this paper, we expand on challenges that are specific to learning-based CC, namely (1) the machine-generated nature of the CC policy which often results in uninterpretable CC behaviour, (2) wild heterogeneity in the implementation of RL-based CC, and, (3) evaluating proposals within a single testbed, while controlling trade-offs related to reproducibility, fidelity, and representativeness of results.

\noindent\textbf{Uninterpretable CC behaviour}. In contrast to human-generated CC policies, RL-based ones are inherently black-boxed. Traditional CC behaviour is interpretable and substantial research has been done to study CC analytically \cite{hespanha2001hybrid, shorten2007modelling} and experimentally \cite{li2007experimental, bbr_analysis}. RL-based CC is driven by an underlying policy (deterministic or stochastic) that is learnt by interacting with a real, emulated, or simulated network, or pre-existing captured traces. The decision-making process that drives the evolution of the transmission rate (and/or congestion window) is embedded within intricate and uninterpretable parametric models. Experimentally studying such RL-based CC must be done extremely carefully; one can identify patterns and provide plausible explanations, but, at the same time, one must be wary of the black-box nature of the underlying CC policy. For example, in Orca the authors argue that fairness arises due to the inclusion of the power metric and additive-increase multiplicative-decrease (AIMD) in decision making \cite{abbasloo2020classic}. However, in Section \ref{fairness}, we show a plethora of experimental setups where Orca is extremely unfair, which puts the original fairness arguments in question. Similarly, Astraea appears to perform exceptionally well in terms of convergence, as shown in \cite{astraea}. However, in Section \ref{convergence} we show that Astraea's convergence profile is affected by fair queue configuration, which may have been part of the experimentation in \cite{astraea}.

\noindent\textbf{Heterogeneity in Implementations}. As there is no agreed mechanism (nor any standardised operating system support, e.g., by exposing a CC API to a learning framework) to implement learning-based CC, existing proposals are based on ad-hoc prototypes implemented as simulation models, user-space applications, kernel modules, or combinations of these. For example, Aurora \cite{jay2019deep} is trained within a bespoke simulator and the learnt model is integrated into a user-space prototype built with UDT \cite{gu2007udt}. Orca, Sage and Astraea on the other hand, are built within the kernel, combined with a user-space component that is performing learning and inference, each leveraging an individual inter-process communication scheme. Sage learns through pre-collected traces of human-derived CC algorithms, and performs inference similarly to Orca. Comparing learning-based CC approaches in a meaningful and reproducible way is far from trivial. 

\noindent\textbf{Reproducibility, Fidelity and Flexibility}. Experimental evaluation of RL-based CC approaches involves several trade-offs. Simulation-based evaluation enables reproducible results and flexibility in designing a wide variety of experiments. However, fidelity of results is degraded substantially, even when state-of-the-art packet-level, discrete event simulators, such as ns-3 or OMNeT++, are used. `In the wild' experimentation enables high fidelity, however, reproducibility is problematic because several parameters (e.g., network topology, routing, background traffic) may be unknown or out of control, when conducting an experiment. Flexibility is also limited as it is extremely difficult to access network environments that are representative of the variety present in the real world (e.g. LEO satellite networks). Network emulation offers a middle ground for these trade-offs. Fidelity is substantially higher compared to a simulated environment but there are limitations related to CPU overhead, hardware offloading, and attached queuing disciplines. Reproducibility of results is possible if all network parameters are recorded. Nondeterminism associated with running an RL scheme on a real system can result in different outcomes, so it is important that multiple runs per experiment are included to assess the level of variability for each measured metric. Network emulation provides substantial flexibility in experimentation, but this needs to be balanced with fidelity-related constraints.

%% file: sections/Methodology.tex
\section{Methodology}
\label{methodology}

In this study, we adopt an emulation-based approach for two reasons; (1) seminal work on learning-based CC has yielded open-source prototypes (see Table \ref{table:rltable} in Section \ref{related_work}) that can be tested on real-world and emulated networks; (2) we aim to maximise fidelity, reproducibility, and flexibility, therefore `in the wild' experimentation was deemed unsuitable.

\noindent\textbf{Selected CC Approaches}. In Section \ref{related_work}, we briefly discuss all different RL-based CC approaches, categorising them into clean slate and hybrid ones. In this study, we expand on both the learning-based and baseline algorithms, compared with our earlier work in \cite{10621288}. More specifically, we include Orca \cite{abbasloo2020classic}, Sage \cite{sage} and Astraea \cite{astraea}, as RL-based CC algorithms, and PCC Vivace \cite{pccvivace}, as an interpretable learning rate-control algorithm. We compare these approaches with Cubic and BBRv3, as they involve both loss-based (Cubic and BBRv3) and delay-based (BBRv3) traits.\footnote{BBRv3 has not been studied much beyond \cite{promisesofbbrv3, bbrv3wired}, therefore, this study also contributes in better understanding its behaviour in a wide range of network parameters and single- and multi-bottleneck topologies. Since our main focus is on learning-based CC, we omit profiling the myriad of TCP CC variants available in the literature.} We have dropped Aurora from our experimentation to maintain legibility in our plots. We have studied Aurora extensively in \cite{10621288}, where it became apparent that its performance is poor (also shown in \cite{astraea}) in real-world network settings, particularly when multiple flows compete with each other or TCP flows. Unfortunately, although we would have liked to include more recent work (e.g., as in \cite{jury, mutant}), the source code for these, or a clear procedure to reproduce the published results, is not available. For Orca, Sage and Astraea, we employ the pre-trained models provided by the authors. Finally, for PCC Vivace, we use its user-space implementation because the kernel-based one generated inconsistent results that could not be attributed to Vivace's key characteristics. 

\noindent\textbf{Performance Metrics}. As the implementations of the selected CC approaches are different from each other, not all performance metrics are collected in the same fashion for all approaches. \textit{Application goodput} was measured differently between them; for Orca, Astraea and Sage, goodput measurement was embedded in the receiver included in the authors' source code. Vivace's goodput was measured using the UDT \cite{gu2007udt} data collection API. For Cubic and BBRv3, we used \textit{iPerf3} \cite{iperf3} to measure goodput. For the \textit{round trip time} (RTT) and \textit{congestion window}, we used \textit{socket statistics} \cite{hemminger2011tcp} to log the values directly from the socket state for BBRv3, Cubic, Orca and Sage. For Astraea, we recorded the RTT and congestion window through its client application. For Vivace, we used the UDT API to measure the RTT and sending rate; Vivace does not have a congestion window. To record \textit{retransmissions} and \textit{link utilisation}, we used \textit{sysstat} \cite{godard2015sysstat}.

\begin{figure*}[t]
    \centering
    \includegraphics[width=0.9\textwidth]{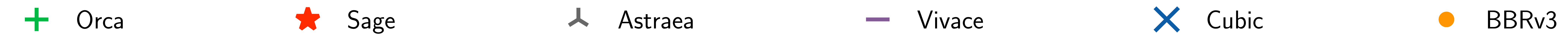}
    \begin{subfigure}[b]{0.3\textwidth}
        \centering
        \includegraphics[width=\textwidth]{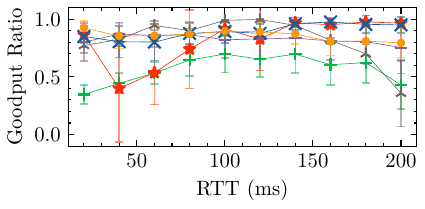}
        \caption{Buffer Size: $0.2\times$ BDP}
        \label{fig:intra_0.2}
    \end{subfigure}
    \hfill
    \begin{subfigure}[b]{0.3\textwidth}
        \centering
        \includegraphics[width=\textwidth]{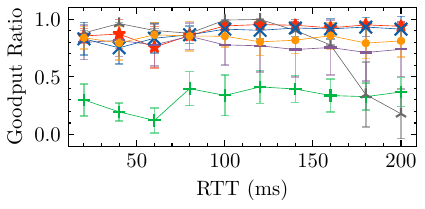}
        \caption{Buffer Size: $1\times$ BDP}
        \label{fig:intra_1}
    \end{subfigure}
    \hfill
    \begin{subfigure}[b]{0.3\textwidth}
        \centering
        \includegraphics[width=\textwidth]{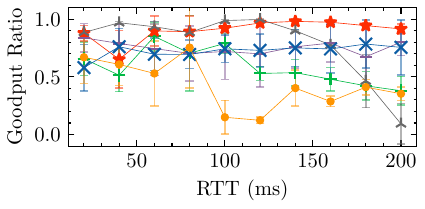}
        \caption{Buffer Size: $4\times$ BDP}
        \label{fig:intra_4}
    \end{subfigure}
    \caption{\textbf{Intra-RTT Fairness}. Goodput ratio for two competing flows in a dumbbell topology. Bottleneck capacity is 100Mbps, both flows experience the same base RTT (shown on x-axis), buffer capacity is set to $0.2\times$ (a), $1\times$ (b), and $4\times$ (c) the BDP.}
\label{fig:intra}
\end{figure*}

\noindent\textbf{Experimental Setup}. Our emulation environment is built using \textit{Mininet} \cite{mininet}. All network nodes (hosts and routers) are represented as network namespaces in our Linux-based emulation host with the following specs: AMD Ryzen Threadripper PRO 7965WX 24 core processor, 64 GB of memory, running Ubuntu 22.04 LTS using the BBRv3 enabled 6.4.0 kernel provided by \cite{bbrthreekernel}, patched with the kernel components required for Astraea and Orca to work. Sage authors only provide a pre-patched and pre-compiled kernel (version 4.19) which we were unable to run on bare metal due to hardware compatibility issues. We therefore have run Sage separately on top of a Hyper-V hypervisor to incur the least overhead.\footnote{We are confident that the virtualisation overhead does not affect Sage's performance because we have also run all other approaches on top of the hypervisor and the results were consistent with the ones produced when run on bare metal and presented here.} 

Experimentation is performed in emulated dumbbell and parking lot topologies. We used \textit{netem} to emulate propagation delay and a token bucket filter to regulate transmission rate at bottlenecks. All other network interfaces are deployed without any propagation delay or rate limiter in place. Hosts are configured to be senders or receivers in only a single data flow at any given time, and we set their send and receive buffers to a large value so that the bottleneck is always in the network (and not on the sending host or as a result of flow control) for all the different buffer capacities tested in this paper. Queuing at the bottleneck interface is tail-drop. To avoid emulation artifacts related to hardware offloading (e.g., TCP segment offloading is performed when forwarding segments, adding unnecessary latency), we have disabled hardware offloading altogether and experimentally verified that the resulting induced CPU overhead does not affect our experiments. Each experiment is run five times, and we report average values and standard deviation in error bars or shaded areas.

%% file: sections/Results.tex
\section{Experimental Evaluation}
\label{evaluation}

In this section, we present the results of our $404$-hour long experiment during which $6270$ Orca, Sage, Astraea, Vivace, Cubic and BBRv3 flows were analysed. We have collected metrics related to network interfaces (e.g., utilisation, retransmissions), CPU and memory parameters (e.g., CPU load and memory usage) and the transport layer (e.g., congestion window, round trip time). The codebase and configuration files required to reproduce the results, and the extracted dataset are available on \textit{figshare} \cite{mazilu_giacomoni_parisis_2025_data} and \cite{mazilu_giacomoni_parisis_2025_code}.

\subsection{Fairness}
\label{fairness}

Fairness in sharing a bottleneck's bandwidth is a very important property that CC algorithms must have. AIMD CC has been shown to yield a fair allocation of bandwidth to competing flows with the same base RTT, but it is problematic when competing flows experience different base RTTs. To date, little is known about the fairness of learning-based CC; here, we focus on the effect that the base RTT, available bandwidth, and buffer capacity have on fairness when two flows contend for bandwidth. The first flow runs for $2000\times$ its \textit{base RTT} (i.e., the delay a packet experiences when the network is empty); the second flow starts after the first one has been running for $25$\% of its full duration.

\subsubsection{Fairness and RTT Variation}
\label{fairness_rtt}

First, we examine how the base RTT that flows experience affects the fairness profiles of the selected CC approaches. We differentiate between the \textit{intra-RTT} and \textit{inter-RTT} cases, where flows experience the same or different base RTT values, respectively. The bottleneck bandwidth is set to $100$Mbps.

\begin{figure*}[ht]
    \centering
    \includegraphics[width=1\textwidth]{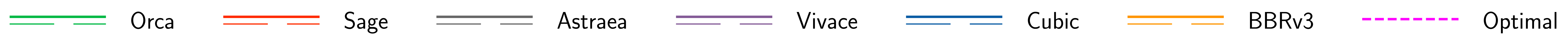}
    \begin{subfigure}[t]{0.325\textwidth}
        \centering
        \includegraphics[width=\textwidth]{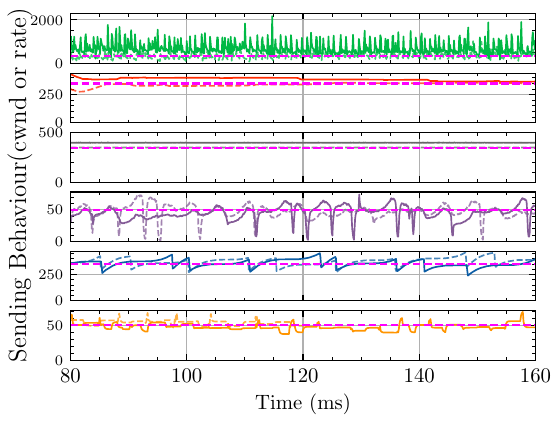}
        \caption{RTT: 80ms, Buffer Size: $0.2\times$ BDP}
        \label{fig:cwnd_80_0.2}
    \end{subfigure}
    \hfill
    \begin{subfigure}[t]{0.325\textwidth}
        \centering
        \includegraphics[width=\textwidth]{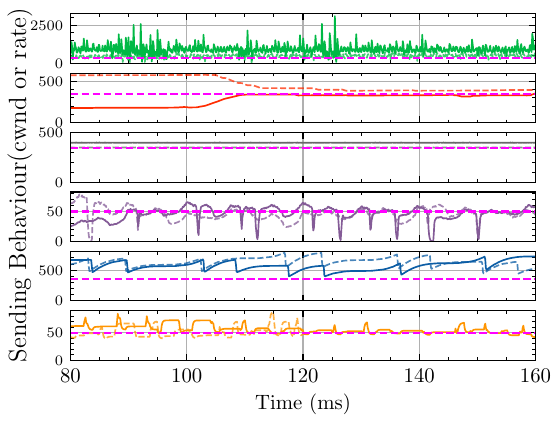}
        \caption{RTT: 80ms, Buffer Size: $1\times$ BDP}
        \label{fig:cwnd_80_1}
    \end{subfigure}
    \hfill
    \begin{subfigure}[t]{0.325\textwidth}
        \centering
        \includegraphics[width=\textwidth]{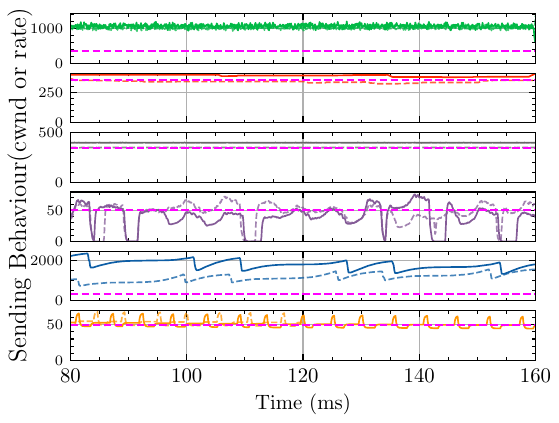}
        \caption{RTT: 80ms, Buffer Size: $4\times$ BDP}
        \label{fig:cwnd_80_4}
    \end{subfigure}
    \caption{\textbf{Intra-RTT Fairness}. Congestion window (sending rate for Vivace and BBRv3) for two competing flows in a dumbbell topology. Bottleneck capacity is $100$Mbps, base RTT is $80$ms, buffer capacity is set to $0.2\times$, $1\times$ and $4\times$ the BDP.}
    \label{fig:cwnd_20_80}
\end{figure*}

\begin{figure*}[ht]
     \centering
     \includegraphics[width=0.9\textwidth]{plots/protocol_legend_markers.pdf}
     \begin{subfigure}[b]{0.3\textwidth}
         \centering
         \includegraphics[width=\textwidth]{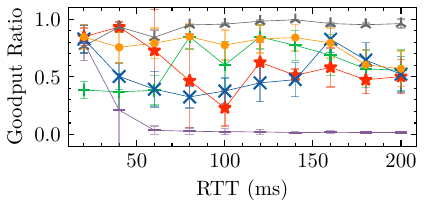}
         \caption{Buffer Size: $0.2\times$ BDP}
         \label{fig:inter_0.2}
     \end{subfigure}
     \hfill
     \begin{subfigure}[b]{0.3\textwidth}
         \centering
         \includegraphics[width=\textwidth]{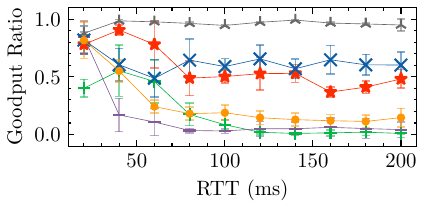}
         \caption{Buffer Size: $1\times$ BDP}
         \label{fig:inter_1}
     \end{subfigure}
      \hfill
     \begin{subfigure}[b]{0.3\textwidth}
         \centering
         \includegraphics[width=\textwidth]{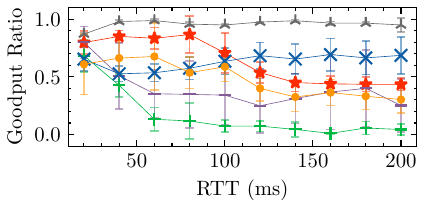}
         \caption{Buffer Size: $4\times$ BDP}
         \label{fig:inter_4}
     \end{subfigure}
     \caption{\textbf{Inter-RTT Fairness}. Goodput ratio for two competing flows in a dumbbell topology. Bottleneck capacity is $100$Mbps and buffer capacity is set to $0.2\times$ (a), $1\times$ (b), and $4\times$ (c) the BDP of the path with the smallest RTT. Flows experience different RTTs; RTT of first flow  is set to $20$ms and RTT of second flow is shown on x-axis.}
     \label{fig:inter}
\end{figure*}

\noindent\textbf{Intra-RTT Fairness}. Figures \ref{fig:intra}a-c show the goodput ratio of the flows for when the buffer capacity is set to $0.2\times$, $1\times$ and $4\times$ their path's BDP, respectively. Flows experience the same base RTT, reported on the x-axis. We report results for the last $500$ RTTs of the experiment so that all CC schemes have the chance to converge to a stable bandwidth allocation.

Orca promises fairness by (1) integrating power in its reward function and (2) AIMD in its operation. Orca yields a substantially more unfair allocation of bandwidth compared to Cubic when the buffer capacity is set to $1\times$ the BDP (Figure \ref{fig:intra_1}), with goodput ratios consistently under 0.5; i.e., one flow grabbing at least more than double the bandwidth allocated to the other one. When the queue size is $0.2\times$ the BDP, Orca performs much better, yet it is still more unfair than Cubic. With the buffer capacity set to $4\times$ the BDP (Figure \ref{fig:intra_4}), Orca’s fairness improves, but it outperforms Cubic only for some base RTT values, remaining less fair otherwise.
Sage maintains high fairness for most tested queue sizes. Sage is trained using a pool of policies, split across two sets containing traces of existing schemes found in the Linux kernel (including BBRv2, Cubic and Vegas). Set 1 encompasses single flow scenarios with changing network conditions. Set 2 is aimed at providing Sage with observations of TCP friendliness. The schemes used to train Sage are effective when it comes to intra-RTT fairness and, expectedly, Sage performs very well in this setup. It is important to note that, while Sage's training bandwidth and base RTT parameter spaces are within the range we use here, the selected queue size of $0.2\times$ the path's BDP is not; this is the likely cause of Sage being less fair (and with high variance) in several of the tested base RTT values (see Figure \ref{fig:intra_0.2}). Astraea, the first RL-based CC scheme to introduce fairness directly in its reward function, in combination with multi-agent RL, maintains high fairness across the parameter space in which it is trained ($10$ to $140$ ms of base RTT). When both of the flows in the experiment are outside the training parameters, we see that fairness tapers off, as seen from the data points in the range of $160$ - $200$ ms in Figure \ref{fig:intra}, a trend that we observe throughout this study. The authors of Vivace provide a theorem, stating that when any number of senders share a bottleneck link and all senders share the Vivace utility function, the sending rates converge to a fair configuration. As we see in Figures \ref{fig:intra}a - \ref{fig:intra}c,  Vivace achieves high fairness (but not higher than TCP Cubic) across the base RTTs for all tested queue sizes. Cubic maintains good fairness properties across the parameter space. When the queue size is $4\times$ the BDP, fairness declines, and this is because Cubic flows converge very slowly to a fair share due to their large BDP values. BBRv3 fairness profiles when the queue size is $0.2\times$ and $1\times$ the path's BDP are very good. When the queue size is $4\times$ the BDP, BBRv3 shows substantial unfairness. Having looked at individual BBRv3 runs, we observed that the second flow exits the start-up phase prematurely. Therefore, it cannot capture a fair share of the bandwidth, because the maximum observed bandwidth, which is used to calculate the path's BDP, is very low. This is because, when the second flow starts immediately after the probe RTT phase of the first flow, it cannot initially capture much bandwidth. Variability in the start time of the second flow, introduced by emulation noise, leads to these timing discrepancies. This trend was also observed in \cite{promisesofbbrv3}.

\noindent\textbf{Congestion Window Evolution}. To better understand fairness, we look at the evolution of the congestion window for competing flows (pacing rate for Vivace and BBRv3). Figure \ref{fig:cwnd_20_80} shows this when the base RTT is $80$ms; a value where Orca performs well when the buffer capacity is $4\times$ the BDP. We plot results for the duration at which flows co-exist. We also plot the line denoting fair allocation that maximises network utilisation and minimises experienced latency; i.e., $\frac{BDP}{2}$ for Orca, Sage Astraea, and Cubic, and $\frac{bandwidth}{2}$ for Vivace and BBRv3. For BBRv3, we show the pacing rate, which is calculated using the \textit{bytes in flight}, rather than the congestion window. This is because BBR purposely uses the congestion window as a limit to the value of bytes in flight, to combat ACK aggregation or compression \cite{cardwell2016bbr}, and therefore the congestion window is a much larger quantity than \textit{bytes in flight}.

Orca’s congestion window progression is noisy particularly in Figures \ref{fig:cwnd_80_0.2} and \ref{fig:cwnd_80_1}. This is because Orca issues coarse-grained congestion window updates at each monitoring interval, while Cubic continuously performing per-ACK window adjustments \cite{abbasloo2020classic}. When the buffer capacity is $0.2\times$ the BDP (Figure \ref{fig:cwnd_80_0.2}), the progression appears to be extremely noisy, with large spikes. When the buffer capacity is set to $1\times$ the BDP (Figure \ref{fig:cwnd_80_1}), values continue to be noisy, oscillating around an unfair allocation. Finally, when the buffer capacity is set to $4\times$ the BDP (Figure \ref{fig:cwnd_80_4}), Orca sets the congestion window to a fair allocation, closer to the optimal allocation compared to Cubic; maintaining high link utilisation and lower latency than Cubic. Sage's congestion window does not fluctuate much after convergence, which is otherwise slow. For example, when the buffer capacity is $1\times$ the BDP (see red lines in Figure \ref{fig:cwnd_80_1}), we see Sage converging to a fair and optimal allocation 110 seconds into the experiment. Looking at Astraea in Figure \ref{fig:cwnd_20_80}, we see how stable and close to optimal the congestion window is, where there is roughly only a 50 packet difference between the congestion windows of the two flows. We have observed this to be the case across all different queue capacities. The sending rate of Vivace fluctuates as it performs its probing trials, which are part of its gradient-ascent learning scheme. Vivace is quite noisy across the different buffer capacity values in Figure \ref{fig:cwnd_20_80}. When the queue size is set to $4\times$ the BDP (Figure \ref{fig:cwnd_80_4}), the reversal in the direction of the sending rate \cite{pccvivace} leads to periodic under-utilisation. Cubic's congestion window evolution follows the known Cubic pattern, which yields a close-to-optimal allocation when the buffer capacity is $0.2\times$ the BDP (Figure \ref{fig:cwnd_80_0.2}). As Cubic is loss-based, it fills the buffer before loss occurs, and this is why the congestion window is much larger than the optimal one for larger buffer capacities (Figures \ref{fig:cwnd_80_1} and \ref{fig:cwnd_80_4}). The pacing rate of BBRv3 follows the optimal threshold well. In Figure \ref{fig:cwnd_80_1} we observe BBRv3 initially overestimating the bottleneck bandwidth; then converging to a fairer allocation. Note that the probe RTT phase, with the characteristic drop in the number of segments sent in the network, is not visible in Figure \ref{fig:cwnd_20_80}. This is because BBRv3 limits its sending rate in the probe RTT phase by decreasing the congestion window, and not the pacing rate (which is what we plot here).

\begin{figure*}[ht]
    \centering
    \includegraphics[width=0.90\textwidth]{plots/protocol_legend_markers.pdf}
    \begin{subfigure}[b]{0.3\textwidth}
        \centering
        \includegraphics[width=\textwidth]{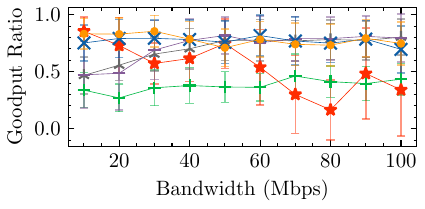}
        \caption{Buffer Size: $0.2\times$ BDP}
        \label{fig:intra_bw_0.2}
    \end{subfigure}
    \hfill
    \begin{subfigure}[b]{0.3\textwidth}
        \centering
        \includegraphics[width=\textwidth]{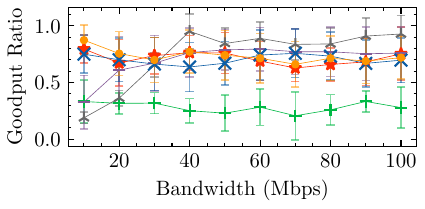}
        \caption{Buffer Size: $1\times$ BDP}
        \label{fig:intra_bw_1}
    \end{subfigure}
    \hfill
    \begin{subfigure}[b]{0.3\textwidth}
        \centering
        \includegraphics[width=\textwidth]{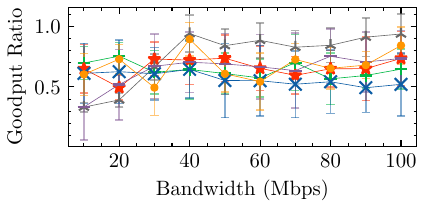}
        \caption{Buffer Size: $4\times$ BDP}
        \label{fig:intra_bw_4}
    \end{subfigure}
    \setlength{\belowcaptionskip}{-10pt}
    \caption{\textbf{Fairness with Bandwidth Variation}. Goodput ratio for two competing flows in a dumbbell topology. Base RTT is $40$ms, bottleneck bandwidth varies as shown on the x-axis, buffer capacity is set to $0.2\times$ (a), $1\times$ (b), and $4\times$ (c) the BDP.}
    \label{fig:intra_bw}
\end{figure*}

\noindent\textbf{Inter-RTT Fairness}. To assess fairness when competing flows have different base RTTs, we repeat the experiment described above but fix the base RTT of the first flow to $20$ms and vary that of the second from $20$ms to $200$ms. In Figure \ref{fig:inter}, we report results for the last $500$ RTTs of the experiment to let CC converge to a fair allocation. 

Orca is generally fairer than Cubic, when the buffer capacity is set to $0.2\times$ the BDP of the flows' path, more so as the difference between the base RTTs of the competing flows increases (see Figure \ref{fig:inter_0.2}, from 60ms upwards). For larger buffer capacities ($1\times$ and $4\times$ the BDP), Orca shows persistent substantial unfairness, which becomes more and more pronounced as the difference between the RTTs of the two flows increases (Figures \ref{fig:inter_1} and \ref{fig:inter_4}). Sage does not manage to consistently outperform Cubic, as none of the traces used in training include inter-RTT scenarios. Furthermore, the CC schemes used to train Sage lack heuristics that enable learning fairness in inter-RTT scenarios. In the lower range of base RTT values in Figure \ref{fig:inter}, we see that Sage is fairer than Cubic, but as the RTT increases beyond $100$ms, Cubic performs better. Astraea performs the best among all tested CC approaches, achieving close to optimal goodput ratios across all tested RTTs and queue sizes. Astraea maintains high fairness even when one flow experiences RTT outside its training range ($160$ - $200$ ms in Figure \ref{fig:inter}). Astraea employs an RTT-agnostic reward component to enforce fairness, which, in combination with a fixed monitoring time period (independent of any RTT values), appears to be very effective, as shown in our results. In \cite{pccvivace}, it is shown that when multiple Vivace senders compete over a single bottleneck link, their rates converge to the same fair value. However, it is not clear if the underlying assumption is that all flows experience the same RTT.\footnote{The proof document cited in \cite{pccvivace} is unfortunately inaccessible and there is no inter-RTT experimentation in the paper.} From our results, it becomes evident that Vivace is unfair when competing flows experience different base RTTs, for all tested queue sizes (see Figures \ref{fig:inter_0.2} - \ref{fig:inter_4}). Looking at individual runs, we discovered that the flow with the lower base RTT is the one dominating the bottleneck, and this is likely due to the fact that it is going through the gradient ascent cycle more often. In Vivace, the frequency of this cycle is based on the monitoring interval which is set to 1 RTT. Cubic is known to be unfair when competing flows experience different base RTTs \cite{cubic_paper} and this is evident in Figure \ref{fig:inter}, particularly when the buffer capacity is $0.2\times$ the BDP (Figure \ref{fig:inter_0.2}). BBRv3’s inter-RTT fairness varies greatly with buffer size. When the queue capacity is set to $1\times$ BDP (\autoref{fig:inter_1}), the flow with the larger base RTT sustains a higher in-flight byte volume and starves the other flow. Increasing the queue capacity (see \autoref{fig:inter_4}) results in BBRv3 being fairer. This is because the pacing gain increases from 0.75 to 0.91 during the bandwidth probe down phase \cite{bbrv3slides} and this makes the existing flow yield less of its bandwidth to the joining flow. When the buffer is very small (Figure \ref{fig:inter_0.2}), the flow experiencing the higher base RTT cannot claim a substantially higher buffer occupancy than the $20$ms flow, resulting in a fairer allocation than when the buffer capacity is $1\times$ the path's BDP.

\begin{figure*}[t]
    \centering
    \includegraphics[width=1\textwidth]{plots/protocol_legend_markers.pdf}
    \begin{subfigure}[b]{0.3\textwidth}
        \centering
        \includegraphics[width=\textwidth]{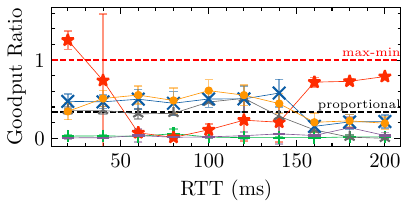}
        \caption{Buffer Size: $0.2\times$ BDP}
        \label{fig:parking_lot_0.2}
    \end{subfigure}
    \hfill
    \begin{subfigure}[b]{0.3\textwidth}
        \centering
        \includegraphics[width=\textwidth]{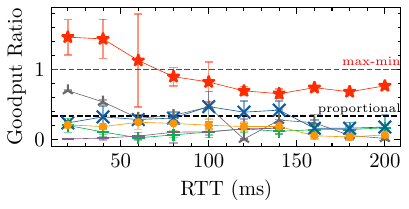}
        \caption{Buffer Size: $1\times$ BDP}
        \label{fig:parking_lot_1}
    \end{subfigure}
    \hfill
    \begin{subfigure}[b]{0.3\textwidth}
        \centering
        \includegraphics[width=\textwidth]{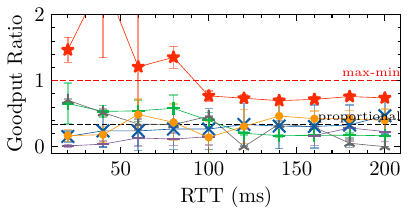}
        \caption{Buffer Size: $4\times$ BDP}
        \label{fig:parking_lot_4}
    \end{subfigure}
    \caption{\textbf{Fairness in a Parking Lot Topology}. Bottleneck capacity is $100$Mbps, all 4 flows experience the same base RTT (shown on x-axis), buffer capacity is set to $0.2\times$ (a), $1\times$ (b), and $4\times$ (c) the BDP.}
    \label{fig:parking_lot}
\end{figure*}

\subsubsection{Fairness and bandwidth variation}
\label{fairness_bandwidth}
We repeat the experiment above using the same setup, but we vary the provisioned bandwidth and fix the base RTT value to $40$ms. Figure \ref{fig:intra_bw} shows the measured goodput ratio for the three different buffer capacity values studied above. When the buffer capacity is set to $0.2\times$ (Figure \ref{fig:intra_bw_0.2}) and $1\times$ (Figure \ref{fig:intra_bw_1}) the BDP, Orca is consistently less fair than Cubic. When the buffer capacity is large (Figure \ref{fig:intra_bw_4}), Orca is fairer to Cubic for several values of the tested bandwidth. For Sage, this experimental setup is not one where the Linux implemented schemes included in its training pool struggle, so it performs well, unable though to meaningfully outperform Cubic. Astraea performs the best across all schemes within its training parameters ($40$ to $100$ Mbps); however, it cannot generalise to bandwidths outside its training range, resulting in a less fair profile. Its fairness remains the highest even when the queue size is $4\times$ the BDP as in Figure \ref{fig:intra_bw_4}. Vivace produces an unfair profile for low bandwidth values, as evidenced in Figure \ref{fig:intra_bw} (particularly \ref{fig:intra_bw_1} and \ref{fig:intra_bw_4}). We believe this is because for such low values, Vivace's probing approach can be very noisy. This is supported by the high variance observed for those data points. Vivace converges to fairer allocations for higher bandwidth values. Cubic and BBRv3 both converge to relatively fair allocations when the buffer capacity is set to $0.2\times$ and $1\times$ the BDP. When this is set to $4\times$ the BDP (Figure \ref{fig:intra_bw_4}), both yield less fair allocations. For Cubic, this is because the higher queue size translates to sparser loss events, leading to Cubic flows taking much longer to converge to a fair allocation. BBRv3 converges much slower as queue sizes increase.

\subsubsection{Fairness in a Multi-Bottleneck Topology}
\label{parking_lot}

In this section, we explore how the different CC schemes behave in a multi-bottleneck setup, which RL-based schemes have not encountered during training. More broadly, such multi-bottleneck scenarios bring a certain level of realism that is otherwise abstracted away when experimenting using dumbbell topologies, and it is important to see how all tested CC schemes perform in terms of fairness. We emulate a \textit{parking lot} topology with three bottlenecks, and concurrently start three flows that cross one of those bottlenecks each. In the experiment, a flow that crosses all bottlenecks is initiated $500$ RTTs. Bottleneck bandwidth is set to $100$Mbps and all flows experience the same base RTT. We plot the average ratio of the goodput of the multi-bottleneck flow over the highest goodput achieved amongst the single-bottleneck flows. We look at the final $500$ RTTs of each run. In Figure \ref{fig:parking_lot}, we plot ratios for all tested base RTT values (shown on the x-axis), for different buffer capacities. We also plot lines for max-min and proportional fairness \cite{fairnesscrowcroft}.

When the buffer capacity is $0.2\times$ the path's BDP (Figure \ref{fig:parking_lot_0.2}), Orca stays below proportional fairness; i.e., the multi-bottleneck flow is dominated by the other three flows. This also holds (but is less prominent) when the buffer capacity is $1\times$ the BDP (Figure \ref{fig:parking_lot_0.2}). When the buffer capacity is $4\times$ the BDP (Figure \ref{fig:parking_lot_4}), Orca comes close to max-min fairness (i.e., all flows get very similar shares of the available bandwidth) for small base RTTs, and close to proportional fairness (i.e., $25$ Mbps for the multi-bottleneck flow and $75$ for each of the single-bottleneck flows) for larger base RTTs. Sage's fairness is closer to max-min fairness as opposed to proportional fairness. In Figures \ref{fig:parking_lot_1} and \ref{fig:parking_lot_4}, the multi-bottleneck Sage flow is very aggressive, particularly for the first few base RTT values ($20$, $40$ and $60$ ms) where it is completely unfair to the single-bottleneck flows (with a ratio that exceeds max-min fairness). When the buffer capacity is small and outside Sage's training range (Figure \ref{fig:parking_lot_0.2}), fairness varies considerably between different base RTT values, indicating poor generalisation. Astraea is the only learning-based CC scheme that stays close to proportional fairness for all buffer capacities (see Figure \ref{fig:parking_lot}), when operating within its training parameters. Astraea explicitly accounts for fairness in its reward function, and this is effective in the multi-bottleneck setup despite the fact that one flow competes for bandwidth separately with three other flows. Vivace does not perform well, and for all different buffer capacity and base RTT values, single-bottleneck flows dominate the multi-bottleneck one. This is because the multi-bottleneck flow, which starts after all single-bottleneck flows do, cannot capture enough bandwidth to come close to proportional fairness. The added latency from crossing three bottleneck buffers results in a behaviour similar to the \textit{Inter-RTT Fairness} case above, despite the fact that all flows experience the same base RTT. Cubic's fairness profile is close to proportional fairness; this has been previously observed for AIMD CC schemes \cite{fairnesscrowcroft}. BBRv3 also comes close to proportional fairness for all tested buffer capacity values.

\begin{figure*}[ht]
    \centering
    \includegraphics[width=0.9\textwidth]{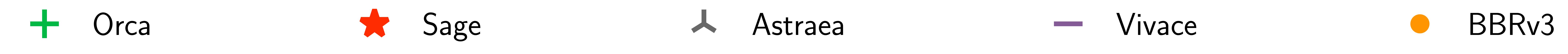}
    \begin{subfigure}[b]{0.3\textwidth}
        \centering
        \includegraphics[width=\textwidth]{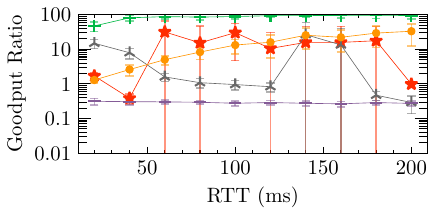}
        \caption{Buffer Size: $0.2\times$ BDP}
        \label{fig:backward_0.2}
    \end{subfigure}
    \hfill
    \begin{subfigure}[b]{0.3\textwidth}
        \centering
        \includegraphics[width=\textwidth]{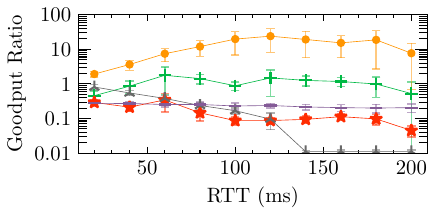}
        \caption{Buffer Size: $1\times$ BDP}
        \label{fig:backward_1}
    \end{subfigure}
    \hfill
    \begin{subfigure}[b]{0.3\textwidth}
        \centering
        \includegraphics[width=\textwidth]{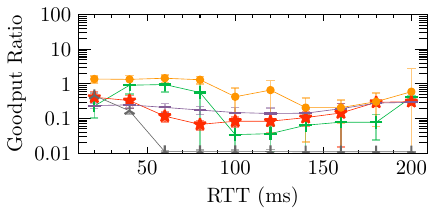}
        \caption{Buffer Size: $4\times$ BDP}
        \label{fig:backward_4}
    \end{subfigure}
    \caption{\textbf{TCP Friendliness}. Goodput ratio for two competing flows (one being Cubic) in a dumbbell topology. Bottleneck capacity is $100$Mbps, flows experience the same base RTT (x-axis), buffer capacity is $0.2\times$ (a), $1\times$ (b), and $4\times$ (c) the BDP.}
    \label{fig:backward}
\end{figure*}

\begin{figure*}[ht]
    \centering
    \includegraphics[width=1\textwidth]{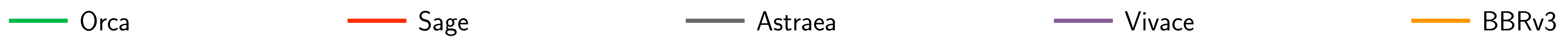}
    \begin{subfigure}[t]{0.325\textwidth}
        \centering
        \includegraphics[width=\textwidth]{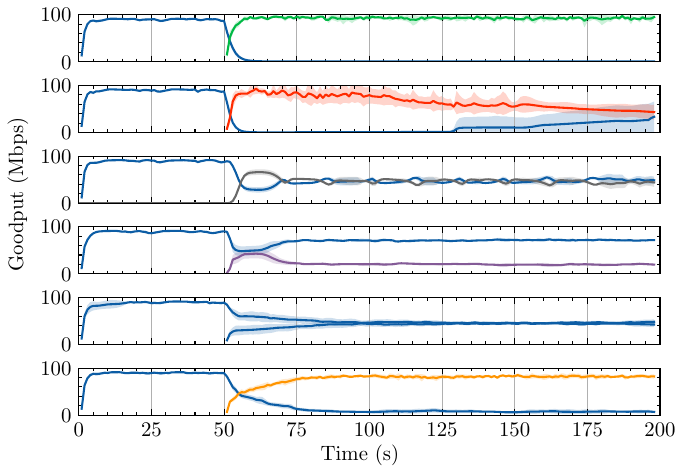}
        \caption{TCP flow first, Buffer Size: $0.2\times$ BDP}
        \label{fig:backward_goodput_0.2_tcp_first}
    \end{subfigure}
    \hfill
    \begin{subfigure}[t]{0.325\textwidth}
        \centering
        \includegraphics[width=\textwidth]{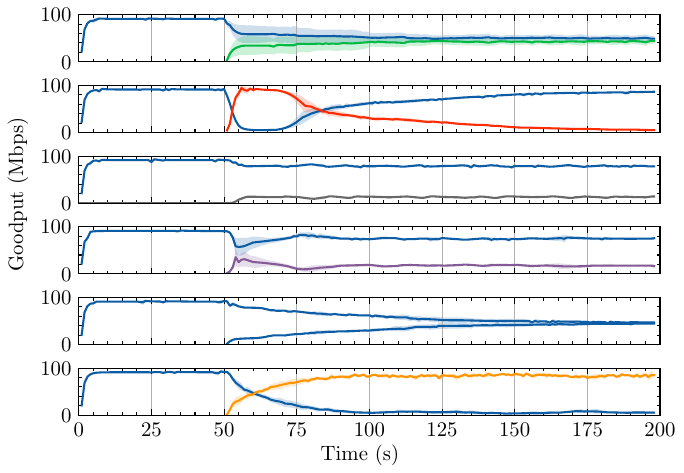}
        \caption{TCP flow first, Buffer Size: $1\times$ BDP}
        \label{fig:backward_goodput_1_tcp_first}
    \end{subfigure}
    \hfill
    \begin{subfigure}[t]{0.325\textwidth}
        \centering
        \includegraphics[width=\textwidth]{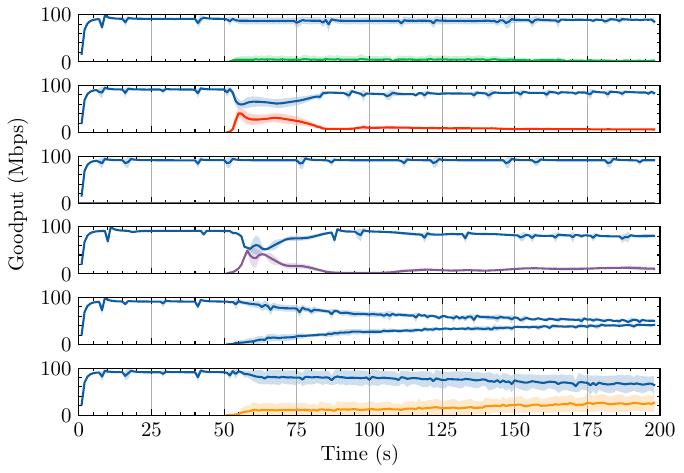}
        \caption{TCP flow first, Buffer Size: $4\times$ BDP}
        \label{fig:backward_goodput_4_tcp_first}
    \end{subfigure}
    \begin{subfigure}[t]{0.325\textwidth}
        \centering
        \includegraphics[width=\textwidth]{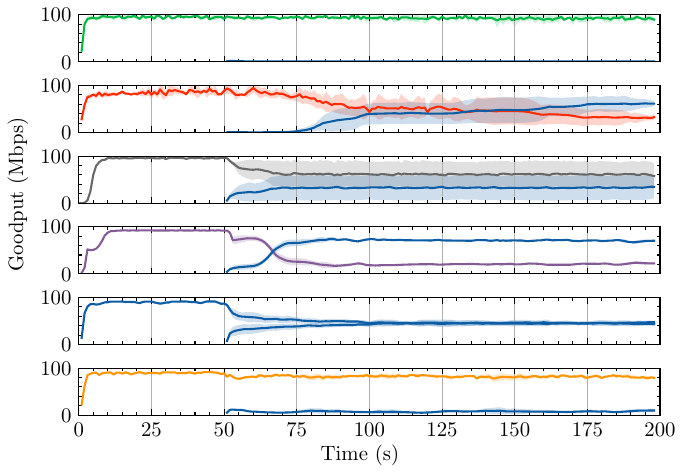}
        \caption{TCP flow last, Buffer Size: $0.2\times$ BDP}
        \label{fig:backward_goodput_0.2_tcp_last}
    \end{subfigure}
    \hfill
    \begin{subfigure}[t]{0.325\textwidth}
        \centering
        \includegraphics[width=\textwidth]{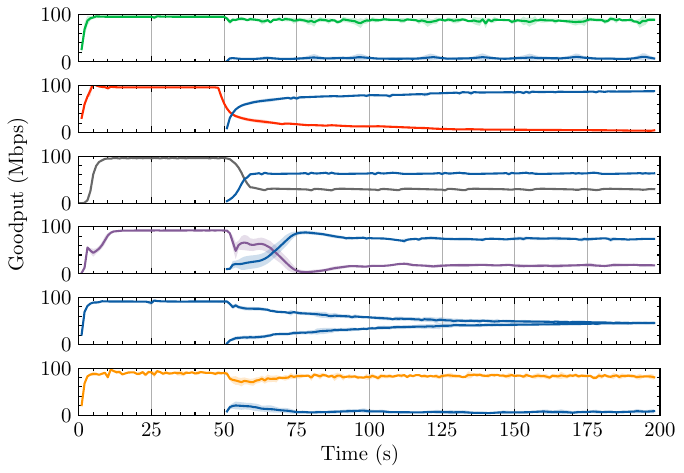}
        \caption{TCP flow last, Buffer Size: $1\times$ BDP}
        \label{fig:backward_goodput_1_tcp_last}
    \end{subfigure}
    \hfill
    \begin{subfigure}[t]{0.325\textwidth}
        \centering
        \includegraphics[width=\textwidth]{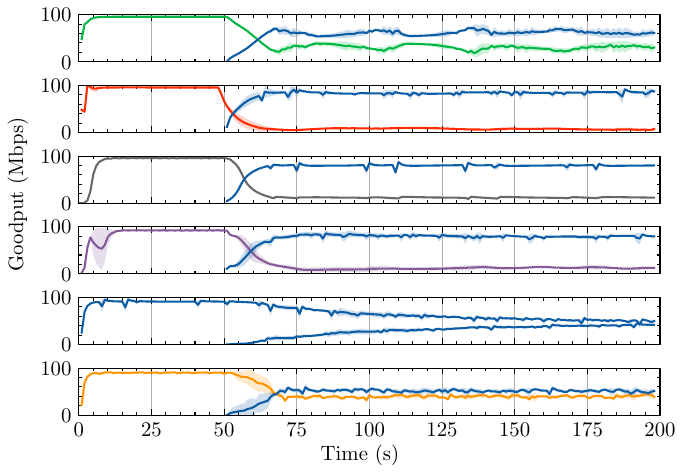}
        \caption{TCP flow last, Buffer Size: $4\times$ BDP}
        \label{fig:backward_goodput_4_tcp_last}
    \end{subfigure}
    \caption{\textbf{TCP Friendliness}. Goodput evolution for two competing flows (one being Cubic) in a dumbbell topology. Bottleneck capacity is $100$Mbps, both flows experience the same base RTT ($100$ms), buffer capacity is set to $0.2\times$, $1\times$, and $4\times$ the BDP.}
    \label{fig:backward_goodput}
\end{figure*}

\subsection{Backward Compatibility}
\label{back_compatibility}

\subsubsection{Two-Flow Setup}

To evaluate the TCP friendliness of learning-based CC, we repeat the fairness experiments but have the first flow running Cubic and the second flow running one of the selected CC algorithms. Figure \ref{fig:backward} shows the goodput ratio of the tested flow over that of the Cubic one during the last $500$ RTTs of the experiment. A scheme is deemed to be friendly to Cubic as long as it does not take up more bandwidth than Cubic, so an allocation can be unfair but TCP friendly. 

When the buffer capacity is set to $0.2\times$ the BDP, Orca is extremely unfair (and unfriendly) when competing with Cubic, with Orca flows dominating Cubic ones, for all base RTT values (Figure \ref{fig:backward_0.2}). When the buffer capacity is $1\times$ the BDP (Figure \ref{fig:backward_1}), Orca is friendlier to Cubic. Finally, when the buffer capacity is $4\times$ the BDP, Orca is friendly to Cubic but the overall allocation is unfair most of the time (Figure \ref{fig:backward_4}). We provide a plausible explanation about this behaviour below when we look at flow dynamics. To address TCP friendliness, Sage employs an additional reward component that measures how fairly each scheme in its training pool shares bandwidth with Cubic \cite{sage}. To showcase its effectiveness, the authors of Sage performed a two-flow experiment under a single parameter instance showing that Sage can fairly share bandwidth with Cubic; we were able to reproduce this specific result. However, in our broader and more systematic experimentation, Sage is shown to achieve an unfair, yet friendly allocation when competing with Cubic, when the buffer capacity is within Sage's training parameter space, where Sage yields most of the bandwidth to the Cubic flow (Figures \ref{fig:backward_1} and \ref{fig:backward_4}). When Sage operates outside its training parameters, it captures more bandwidth than Cubic, with significant variance present in the different runs (Figure \ref{fig:backward_0.2}). Astraea is generally friendly to Cubic, yielding more and more bandwidth to it, as the buffer capacity increases, eventually leading to completely unfair bandwidth allocations (e.g., above $140$ms in Figure \ref{fig:backward_1} and above $60$ms in Figure \ref{fig:backward_4}), where Cubic completely dominates the bottleneck. This is because Astraea's reward function penalises latency and therefore the policy will decrease the congestion window as the Cubic flow fills up the buffer. This delay-induced penalty is much lower (to negligible) when the underlying buffer capacity is very small (Figure \ref{fig:backward_0.2}); as a result, Astraea is friendlier for most of the base RTT values tested, when the buffer capacity is $0.2\times$ the path's BDP with the overall allocation being more fair, also. Vivace is friendly to Cubic across the different buffer capacities, although the overall allocation remains unfair, with Cubic taking substantially more bandwidth (Figures \ref{fig:backward}). With just one Cubic flow competing for bandwidth, the frequency at which the queue gets full is low; therefore, Vivace finds it can reduce RTT and increase its utility by reducing its rate often \cite{pccvivace}; consequently, it achieves lower throughput than Cubic. BBRv3 friendliness to Cubic depends on the buffer capacity. With smaller buffer sizes (Figures \ref{fig:backward_0.2} and \ref{fig:backward_1}), BBRv3 dominates Cubic. When the buffer capacity is set to $4\times$ the BDP and the base RTT values are high (Figure \ref{fig:backward_4}), Cubic fills the buffer, preventing BBRv3 from accurately estimating the bottleneck bandwidth and minimum RTT. As a result, the allocation is friendlier (and fairer) to TCP. It is worth noting that BBRv3 is fairer as the base RTT values get smaller. This is due to the wall clock-based probing mechanism that BBRv2 first introduced to be friendly to loss-based schemes in a parameter space of a typical broadband connection \cite{bbrthreekernel}. 

\begin{figure*}[ht]
    \centering
    \includegraphics[width=0.90\textwidth]{plots/protocol_legend_markers_nocubic.pdf}
    \begin{subfigure}[b]{0.3\textwidth}
        \centering
        \includegraphics[width=\textwidth]{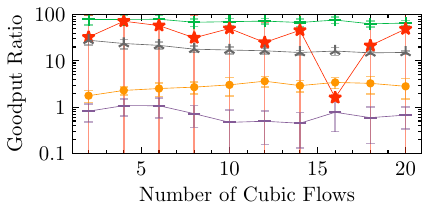}
        \caption{Buffer Size: $0.2\times$ BDP}
        \label{fig:backward_flows_0.2}
    \end{subfigure}
    \hfill
    \begin{subfigure}[b]{0.3\textwidth}
        \centering
        \includegraphics[width=\textwidth]{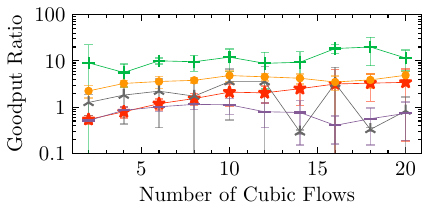}
        \caption{Buffer Size: $1\times$ BDP}
        \label{fig:backward_flows_1}
    \end{subfigure}
    \hfill
    \begin{subfigure}[b]{0.3\textwidth}
        \centering
        \includegraphics[width=\textwidth]{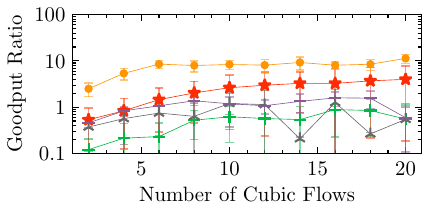}
        \caption{Buffer Size: $4\times$ BDP}
        \label{fig:backward_flows_4}
    \end{subfigure}
    \caption{\textbf{TCP Friendliness}. Goodput ratio between the tested scheme and the average goodput achieved by the Cubic flows. Bottleneck capacity is $100$Mbps, RTT is set to $30$ms, buffer capacity is set to $0.2\times$ (a), $1\times$ (b), and $4\times$ (c) the BDP. The number of joining Cubic flows is shown on the x-axis. }
    \label{fig:backward_flows}
\end{figure*}

\noindent\textbf{Flow Dynamics}. To better understand flow dynamics when competing with Cubic, we pick a single base RTT value ($100$ms) and repeat the experiment above in two variations, where Cubic starts first and last, respectively. In Figure \ref{fig:backward_goodput}, we plot the mean and standard deviation (shaded area) of the measured goodput as the two flows compete for bandwidth. 

When the buffer capacity is set to $0.2\times$ the BDP of the path, we observe that Orca dominates the bottleneck link, even when the Cubic flow starts first (Figures \ref{fig:backward_goodput_0.2_tcp_first} and \ref{fig:backward_goodput_0.2_tcp_last}); the buffer is so shallow that Cubic will face loss often, allowing the more aggressive Orca flow to dominate the bottleneck. When the buffer capacity is set to $1\times$ the BDP of the path, Orca behaves differently when the Cubic flow starts first compared to when it starts second. In the former case (Figures \ref{fig:backward_goodput_1_tcp_first}), the bottleneck link runs at its full capacity when the Orca flow starts, and therefore Orca cannot observe the actual base RTT and maximum bandwidth values, which are crucial for Orca's decision making process. As a result, Orca is less aggressive and friendlier to Cubic, with the allocation converging to a fair equilibrium. When Orca starts first and the buffer capacity is set to $1\times$ the BDP, it completely dominates Cubic (Figure \ref{fig:backward_goodput_1_tcp_last}); having observed the base RTT and maximum bandwidth, Orca does not yield much bandwidth to Cubic, which backs off to its aggressive competitor. When the buffer capacity is set to $4\times$ the BDP and the Orca flow starts second, it cannot capture any bandwidth at all; as explained above, Cubic has already captured all available bandwidth, and Orca cannot observe the minimum RTT and maximum bandwidth values, consequently operating at the observed low point (Figure \ref{fig:backward_goodput_4_tcp_first}). Finally, when the buffer capacity is set to $4\times$ the BDP and the Orca flow starts first, we observe an interesting pattern in the measured goodput (Figure \ref{fig:backward_goodput_4_tcp_last}), with Orca and Cubic periodically approaching and diverging from the fair allocation point. We think this is because Orca's hybrid approach relies on Cubic dynamics on a per-acknowledgment basis and, in this case, Cubic manages to capture some bandwidth due to the buffer capacity being high. Sage, trained using traces from schemes with delay-based CC traits, appears to be less aggressive than (and therefore friendly to) Cubic for all buffer capacity values tested, regardless of the order in which the two competing flows start. When the buffer capacity is set to $1\times$ and $4\times$ the BDP, we observe that Sage eventually gives up all its bandwidth to Cubic (Figures \ref{fig:backward_goodput}b, \ref{fig:backward_goodput}c, \ref{fig:backward_goodput}e, and \ref{fig:backward_goodput}f). Interestingly, this happens even when Sage appears to capture most of the bandwidth during its start-up phase (e.g., as in Figure \ref{fig:backward_goodput_1_tcp_first}). When the buffer capacity is $0.2\times$ the BDP, Sage behaves in the same way, although at a much slower pace (Figures \ref{fig:backward_goodput}a and \ref{fig:backward_goodput}d); we can confirm that Sage yields most of its bandwidth, in both cases, when the experiment is left to run for longer (not shown in this Figure). Astraea's reward function is latency-aware and the underlying agent has learned to reduce the congestion window when this would reduce latency. Therefore, when the buffer capacity is large ($1\times$ and $4\times$ the BDP), Cubic attempts to fill the available buffer and the consequent latency increase deters Astraea from pursuing further bandwidth. Therefore, Cubic dominates the bottleneck, resulting in a friendly yet unfair allocation; this is the case both when the Cubic flow starts first (Figures \ref{fig:backward_goodput_1_tcp_first} and \ref{fig:backward_goodput_4_tcp_first}) or last (Figures \ref{fig:backward_goodput_1_tcp_last} and \ref{fig:backward_goodput_4_tcp_last}). When the buffer capacity is $0.2\times$ the BDP, the latency increase when the buffer gets full is not substantial enough for Astraea to reduce its congestion window. Therefore, it behaves more like a loss-based CC scheme that is friendly to Cubic with the overall allocation being fairer, too. Vivace's behaviour does not depend on which flow starts first. As discussed above, Vivace finds it can reduce its observed RTT and, therefore, increase its utility by reducing its rate often \cite{pccvivace}; consequently, it is dominated by Cubic. For this particular base RTT, BBRv3 is extremely unfair (and unfriendly) to Cubic when the buffer capacity is $0.2\times$ and $1\times$ the BDP. For smaller buffer capacities, BBRv3 is completely unfair to Cubic (Figures \ref{fig:backward_goodput}a, \ref{fig:backward_goodput}b, \ref{fig:backward_goodput}d, \ref{fig:backward_goodput}e). When the buffer capacity is $4\times$ the BDP mark, BBRv3 converges to a fair allocation; at different paces depending on which flow starts first. This confirms the findings in \cite{promisesofbbrv3}.

\begin{figure*}[t]
    \centering
    \includegraphics[width=0.9\textwidth]{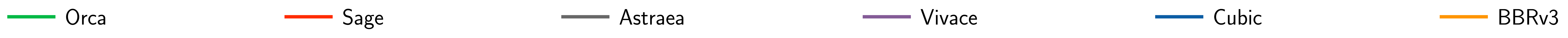}
    \begin{subfigure}[t]{0.325\textwidth}
        \centering
        \includegraphics[width=\textwidth]{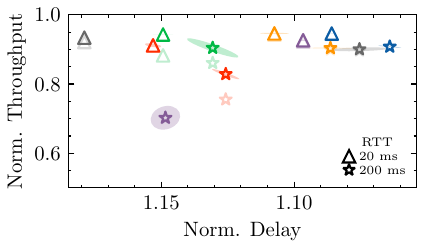}
        \caption{Buffer Size: $0.2\times$ BDP}
        \label{fig:efficiency_0.2}
    \end{subfigure}
    \hfill
    \begin{subfigure}[t]{0.325\textwidth}
        \centering
        \includegraphics[width=\textwidth]{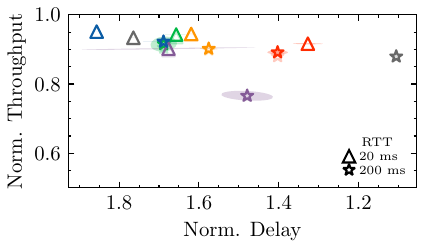}
        \caption{Buffer Size: $1\times$ BDP}
        \label{fig:efficiency_1}
    \end{subfigure}
    \hfill
    \begin{subfigure}[t]{0.325\textwidth}
        \centering
        \includegraphics[width=\textwidth]{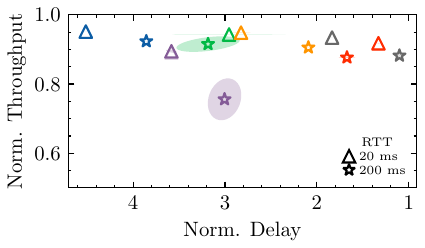}
        \caption{Buffer Size: $4\times$ BDP}
        \label{fig:efficiency_4}
    \end{subfigure}
    \caption{\textbf{Efficiency}. Aggregate network throughput and average latency for four competing flows in a dumbbell topology. Bottleneck capacity is $100$Mbps, flows experience the same base RTT, buffer capacity is set to $0.2\times$, $1\times$ and $4\times$ the BDP.}
    \label{fig:efficiency}
\end{figure*}

\subsubsection{Multi-Flow Setup}

Here, we explore friendliness when a single flow is competing with an increasing number of Cubic flows. Previously, both Astraea and Vivace have been shown to be TCP friendly when competing with many Cubic flows \cite{astraea}\cite{pccvivace}, and so we are interested in reproducing these findings and observing how other schemes perform in the same setup. We start the flow of the selected CC scheme first; then, for the next 500 RTTs, we start Cubic flows with their start times selected uniformly at random within this time period. The base RTT and bottleneck bandwidth are set to $30$ms and $100$Mbps, respectively. In Figure \ref{fig:backward_flows}, we plot the goodput ratio of the tested flow over the average goodput of the Cubic flows; the number of Cubic flows that we start is shown on the x-axis.

Orca's fairness profile when multiple Cubic flows compete with it is similar to the two-flow setup. For example, in Figure \ref{fig:backward_flows_0.2}, the Orca achieves a ratio close to 100, which means that the Orca flow completely starves the incoming Cubic flows, which is what we observed in Figure \ref{fig:backward_0.2}. In Figure \ref{fig:backward_flows_1} we see that Orca yields more bandwidth compared to that in the previous experiment. Similarly, Orca is the friendliest to Cubic when the queue size is $4\times$ BDP (Figure \ref{fig:backward_flows_4}), indicated by the ratio being less than 1. Sage’s friendliness to Cubic when the buffer capacity is $0.2\times$ BDP is poor and extremely variable (Figure \ref{fig:backward_flows_0.2}), with goodput ratio values ranging from 1 to 100. As mentioned in Section \ref{fairness}, this value of the buffer capacity is outside the parameter range used in training Sage. When operating within its training range (Figures \ref{fig:backward_flows_1} and \ref{fig:backward_flows_4}), Sage shares bandwidth with Cubic more fairly and with much lower variance, but as the number of Cubic flows grows, Sage becomes more unfair. Interestingly, these results are very much in line with those presented in \cite{sage}, where the experimental setup involved $3$ and $7$ Cubic flows competing with $1$ Sage flow and it was apparent that Sage becomes more aggressive to Cubic the more Cubic flows are introduced. Although there was no explanation for this behaviour in \cite{sage}, here we conjecture that Sage, having been trained only with setups, including $2$ Cubic flows, it has learnt to be more aggressive the more pressure it gets from its competitor. When the buffer capacity is $0.2\times$ the BDP, Astraea maintains its aggressiveness to Cubic, as observed in the two-flow setup, regardless of the number of Cubic flows; Astraea gets at least $10\times$ the goodput of Cubic flows. When we increase the queue size to $1\times$ the BDP, Astraea is friendlier to Cubic flows (Figure \ref{fig:backward_flows_1}), which confirms the results shown in \cite{astraea}. When the buffer capacity is further increased to $4\times$ the BDP (Figure \ref{fig:backward_flows_4}), Astraea becomes less tolerant to the delay inflation caused by the Cubic flows filling the large buffer, and consequently, yields more bandwidth; the resultant allocation is still friendly but less fair than when the buffer capacity is $1\times$ the BDP. Vivace is friendly to Cubic for the buffer capacity and number of competing Cubic flows explored in this section. This is in line with the trend presented in \cite{pccvivace}\footnote{A direct comparison is not possible as the experimental setup in \cite{pccvivace} uses low bandwidth and allocates buffer proportionally to the number of flows.}, further strengthening the authors' intuition related to the reduced utility gain from the RTT gradient, as also discussed above. BBRv3 generally captures more bandwidth than the average Cubic flow. Specifically, with buffer sizes of $0.2\times$ and $1\times$ the BDP, it captures $2$–$3$ times more bandwidth than an average Cubic flow. When the buffer capacity is increased to $4\times$ the BDP, the unfriendliness is exacerbated and BBRv3 captures bandwidth more aggressively, due to the BBRv3 flow experiencing a much higher base RTT because the larger buffer being filled by the Cubic flows. 

\begin{figure*}[ht]
    \centering
    \includegraphics[width=1\textwidth]{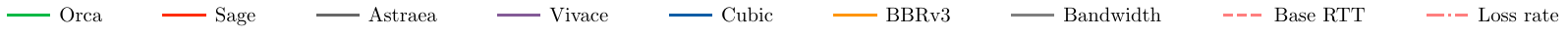}
    \begin{minipage}[t]{0.40\textwidth}
        \centering
        \includegraphics[width=\linewidth]{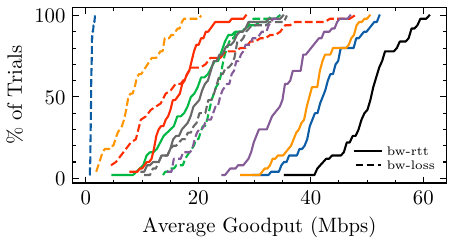}
        \captionsetup{type=figure}
        \caption{\textbf{Responsiveness}: Cumulative Distribution of Goodput}
        \label{fig:response_goodput_cdf}
    \end{minipage}
    \hfill
    \begin{minipage}[t]{0.57\textwidth}
        \centering
        \includegraphics[width=\linewidth]{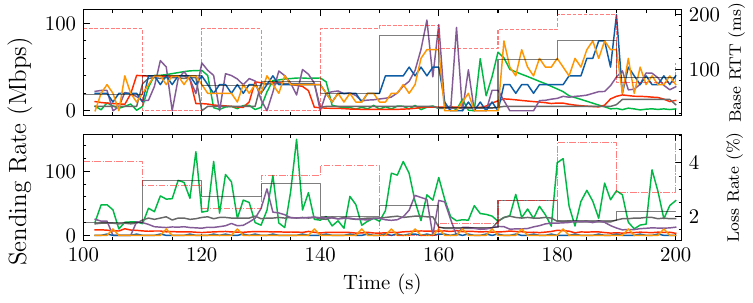}
        \captionsetup{type=figure}
        \caption{\textbf{Responsiveness}: Sending Rate in Time}
        \label{fig:response_send_rate}
    \end{minipage}
\end{figure*}

\subsection{Efficiency}
\label{efficiency}
In this section, we focus on how efficiently the different CC approaches utilise the underlying network resources. We use the same experimental setup as in Section \ref{fairness_rtt}, but with four competing flows and experiment with two base RTT values ($20$ms and $200$ms). Flows are scheduled in $25$-second intervals, each lasting $100$ seconds. We measure aggregate goodput normalised by the capacity of the bottleneck, and average flow latency normalised by the path's base RTT. We also measure the retransmission rate as this affects network utilisation. In Figure \ref{fig:efficiency}, we plot mean values and $1$-$\sigma$ ellipses for the $25$ seconds all flows co-exist. 

When the base RTT is set to $20$ms, all CC schemes appear to perform well with flows capturing most of the available bandwidth. The exception to this is Orca, which performs a bit worse when the buffer capacity is set to $0.2\times$ the BDP. We have observed a large amount of retransmissions from Orca flows in this case, which we attribute to the shallow buffer and the highly aggressive sending rate incurred by Orca. Astraea also incurs retransmissions, but not as significantly as Orca does. When flows experience a $200$ms RTT, the situation is slightly different because the $4$ competing flows converge slower. Consequently, all CC schemes perform worse compared to the $20$ms setup; i.e., for each CC scheme, its star marker is below the triangle one. Vivace is affected the most, as shown in Figure \ref{fig:efficiency}, particularly when the buffer capacity is $0.2\times$ the BDP. As shown in Section \ref{responsiveness}, convergence for Vivace is very noisy with periods where flows severely underperform. Previously, in Figure \ref{fig:cwnd_80_4}, we observed extensive periods of inactivity for Vivace with just $1$ flow running through a bottleneck with a base RTT value of $80$ms. Here, we have $4$ flows and the base RTT is $200$ms and therefore these periods of inactivity because of the reversal in the direction of Vivace's sending rate are much longer. We have confirmed this by looking at individual runs that we do not include here for brevity. Sage's goodput when the capacity is set to $0.2\times$ the BDP is affected by a substantial amount of retransmissions and we attribute that to the fact that it operates outside its training range.

For the latency dimension, we focus on the larger buffer capacity values (i.e., $1\times$ and $4\times$ the BDP) where latency inflation is more apparent and problematic. Orca performs the worst between the RL-based CC approaches, despite including a latency-sensitive component in its reward function. We have previously observed Orca being aggressive in its sending rate and resistant to non-congestive loss, which we believe are the reasons for the latency inflation, despite latency being penalised by its reward function. Sage performs the best in both experimental setups, inflating latency for its flows the least. This is because it has been trained with traces from several latency-aware CC schemes. Astraea also integrates latency in its reward and, in contrast to Orca, this appears to be very effective in keeping latency inflation low, as shown in Figures \ref{fig:efficiency_1} and \ref{fig:efficiency_4}. This is the case for both tested base RTT values but as we show in the next section Astraea suffers from substantial unfairness in the $200$ms setup.

Cubic and BBRv3 incur significant latency inflation for their flows. Cubic just fills up the buffer that is given, as it is a purely loss-based protocol. In \cite{bbrv3wired}, it has been shown that all BBR versions tend to overestimate the bottleneck bandwidth, leading to persistent queue build-up and inflated RTTs. BBRv3 has not introduced mechanisms to explicitly address this issue and is therefore exhibits the same behaviour in this experiment.

\begin{figure*}[ht]
    \centering
    \begin{subfigure}[t]{0.325\textwidth}
        \centering
        \includegraphics[width=\textwidth]{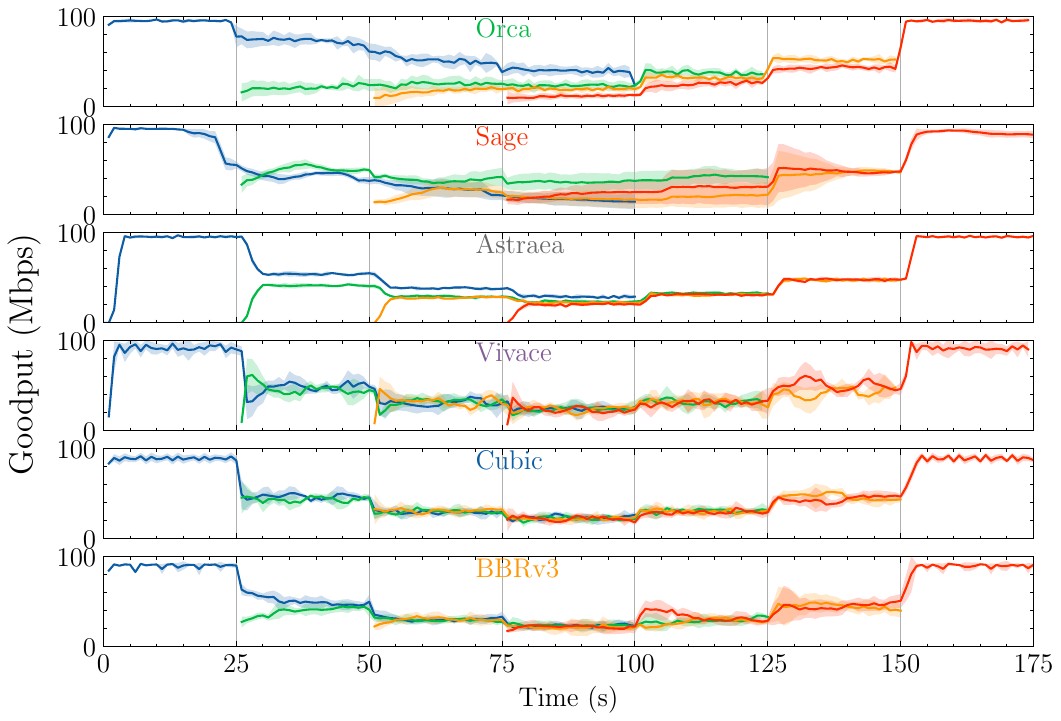}
        \caption{RTT: 20ms, Buffer Size: $0.2\times$ BDP}
        \label{fig:conv_0.2}
    \end{subfigure}
    \hfill
    \begin{subfigure}[t]{0.325\textwidth}
        \centering
        \includegraphics[width=\textwidth]{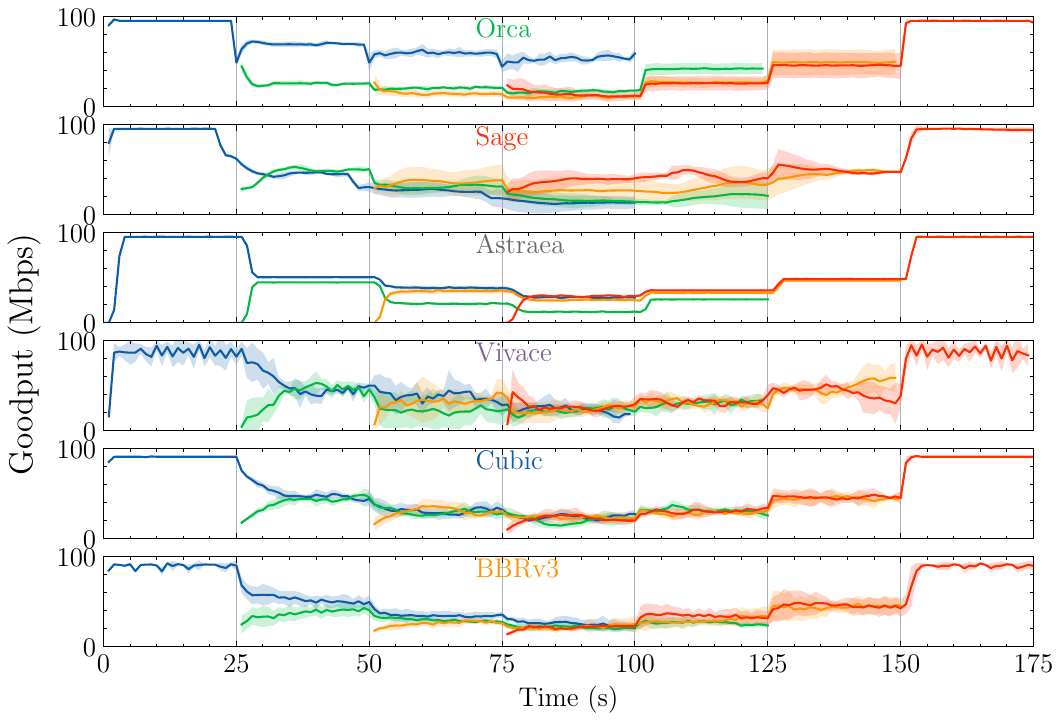}
        \caption{RTT: 20ms, Buffer Size: $1\times$ BDP}
        \label{fig:conv_1}
    \end{subfigure}
    \hfill
    \begin{subfigure}[t]{0.325\textwidth}
        \centering
        \includegraphics[width=\textwidth]{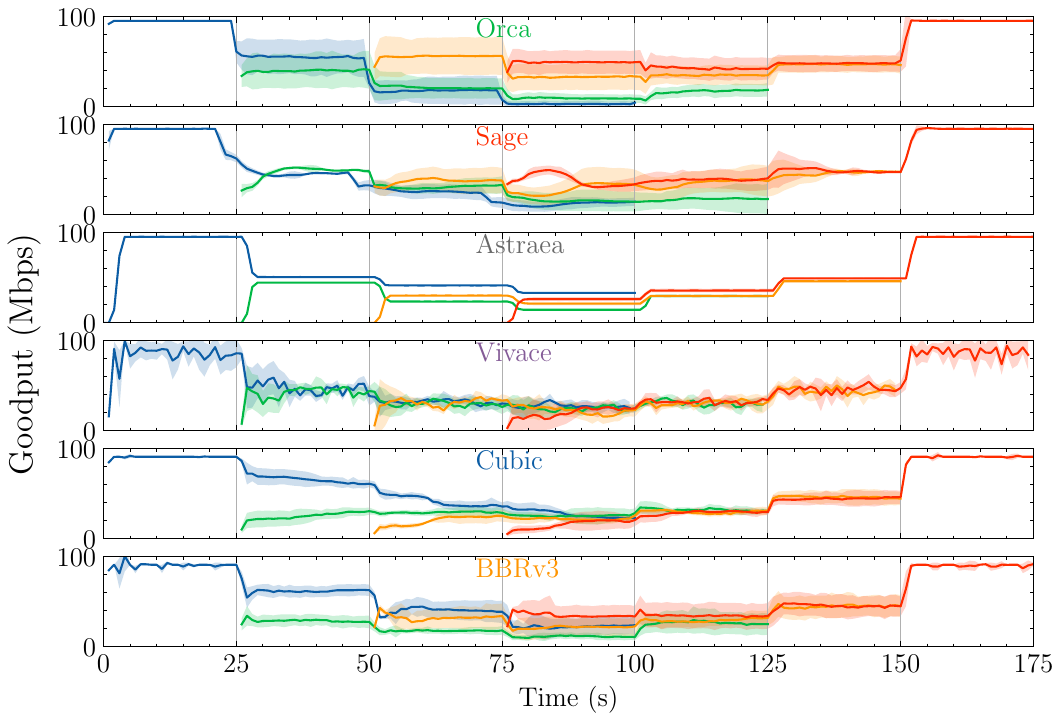}
        \caption{RTT: 20ms, Buffer Size: $4\times$ BDP}
        \label{fig:conv_4}
    \end{subfigure}
    \caption{\textbf{Convergence}. Goodput evolution for four competing flows in a dumbbell topology. Bottleneck capacity is $100$Mbps, flows experience the same base RTT of $20$ms, buffer capacity is set to $0.2\times$, $1\times$ and $4\times$ the BDP.}
    \label{fig:conv}
\end{figure*}

\subsection{Responsiveness}
\label{responsiveness}

Here, we study how the selected CC approaches react to dynamically changing network conditions. Such changes may be occurring because the underlying network changes (e.g., due to link re-configurations and routing changes, or because the topology itself constantly changes, as in low-earth orbit satellite networks) or because of transient hotspots. We examine how CC approaches behave in the presence of bandwidth and base RTT changes, and, separately, bandwidth and non-congestive loss changes.

\noindent\textbf{Varying Bandwidth and Base RTT}. We experiment with a single $300$-second flow in our emulated dumbbell network. Every $10$ seconds we update the bottleneck bandwidth and the base RTT parameters by uniformly selecting values from the ranges $1$ - $100$Mbps and $20$ - $200$ms, respectively. For each run, bandwidth and base RTT parameters are the same for all tested protocols. The buffer capacity is set to $638$KB ($1\times$ the mean BDP). For each CC approach, we repeat the experiment $50$ times and calculate the cumulative distribution of average goodput, as in \cite{pccvivace}. In Figure \ref{fig:response_goodput_cdf}, the results of this experiment are shown in solid lines. In Figure \ref{fig:response_goodput_cdf}, the black solid line denotes the optimal bandwidth.

Orca performs badly, particularly when there are jumps in the base RTT. We attribute this to Orca’s reliance on estimating the minimum RTT and maximum bandwidth which, in these cases, becomes very problematic. This can be clearly seen in Figure \ref{fig:response_send_rate}, at $140$ seconds, where Orca's sending rate collapses when the base RTT increases substantially. Similarly, when the available bandwidth increases, as in Figure \ref{fig:response_send_rate} at $150$ seconds, Orca remains completely unresponsive to the change. Sage performs even worse, frequently not responding to changes in bandwidth and base RTT (as in Figure \ref{fig:response_send_rate} at $140$ and $150$ seconds). The traces that Sage uses for training include changing bandwidth values in its step scenarios \cite{sage}, however, these changes are based on a set of fixed multipliers. In an environment where the bandwidth and base RTT are uniformly selected, as in our experiment in Figure \ref{fig:response_goodput_cdf}, Sage is unable to capture the available bandwidth, achieving the lowest average performance between all CC schemes. Astraea, as with the other RL-based CC schemes, is unable to respond to frequent delay and bandwidth fluctuations, achieving less than half the optimal performance (Figure \ref{fig:response_goodput_cdf}). Note that Astraea is not trained in environments where bandwidth and base RTT change during a training episode. Like Orca, Astraea uses the minimum observed RTT as a baseline and interprets any increase beyond it as a sign of congestion, clearly seen at $120$ seconds in Figure \ref{fig:response_send_rate}. In \cite{pccvivace}, Vivace is shown to outperform all CC schemes it is compared to in a similar responsiveness experiment. In our study, this is not the case, and Vivace performs much worse than Cubic and BBRv3. This is because, in our experimentation, we adopt a much larger delay range ([$20$ms, $200$ms]) than in \cite{pccvivace} ([$10$ms, $100$ms]), which results in a higher average delay experienced by the flow. This, in turn, leads to Vivace being less responsive, because (1) its monitoring interval is based on the underlying observed RTT and (2) Vivace employs a bandwidth growth cap to filter out erroneous values. This phenomenon is clearly shown in Figure \ref{fig:response_send_rate} at 150 seconds, where Vivace is very slow to explore the available bandwidth when the delay jumps to $200$ms. Cubic performs the best with its bandwidth probing heuristics being effective in tracking the available bandwidth well relative to the base RTT it experiences. Similarly, BBRv3 tracks the available bandwidth well. The small deficit compared to Cubic can be attributed to its wall clock bandwidth probes, which cause a slightly slower adaptation to bandwidth increases, as it must wait at most 3 seconds to start the probe bandwidth phase to explore newly available bandwidth. This is clearly shown at $140$ and $180$ seconds, where BBRv3 is slow to adapt to the base RTT increase. 

\begin{figure*}[ht]
    \centering
    \begin{subfigure}[t]{0.325\textwidth}
        \centering
        \includegraphics[width=\textwidth]{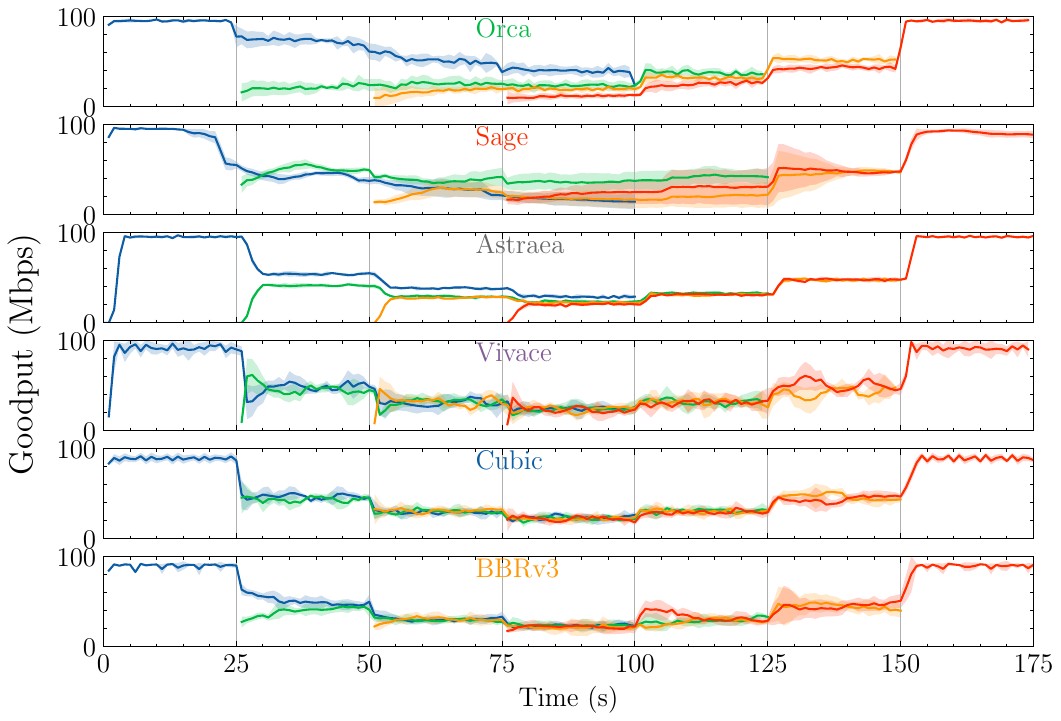}
        \caption{RTT: 200ms, Buffer Size: $0.2\times$ BDP}
        \label{fig:conv_200_0.2}
    \end{subfigure}
    \hfill
    \begin{subfigure}[t]{0.325\textwidth}
        \centering
        \includegraphics[width=\textwidth]{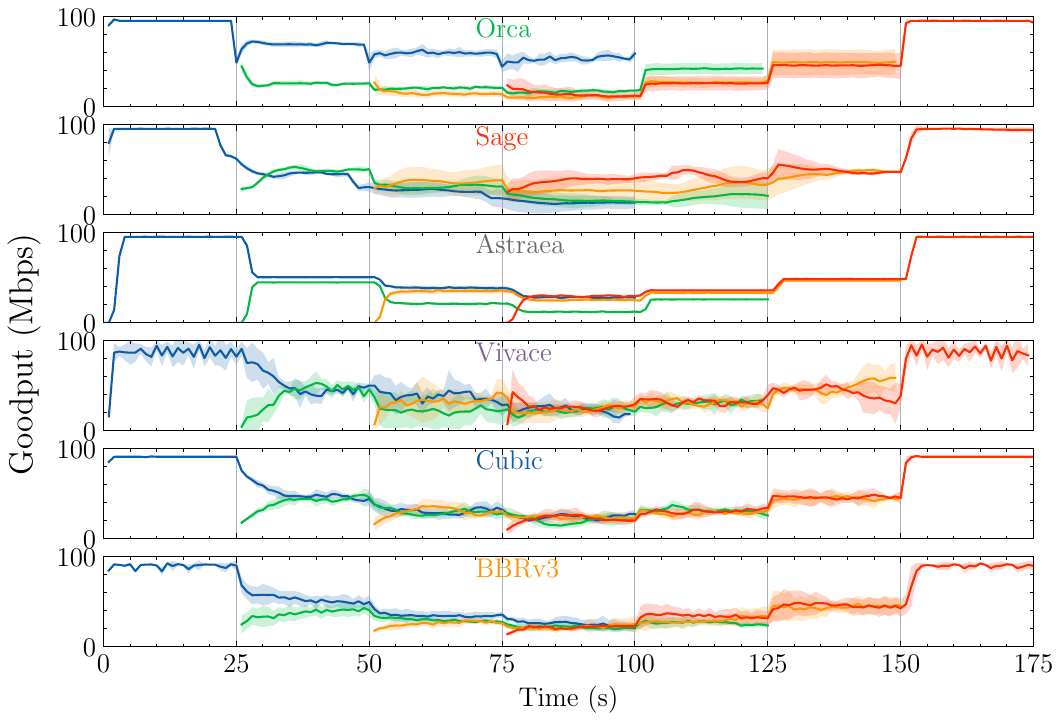}
        \caption{RTT: 200ms, Buffer Size: $1\times$ BDP}
        \label{fig:conv_200_1}
    \end{subfigure}
    \hfill
    \begin{subfigure}[t]{0.325\textwidth}
        \centering
        \includegraphics[width=\textwidth]{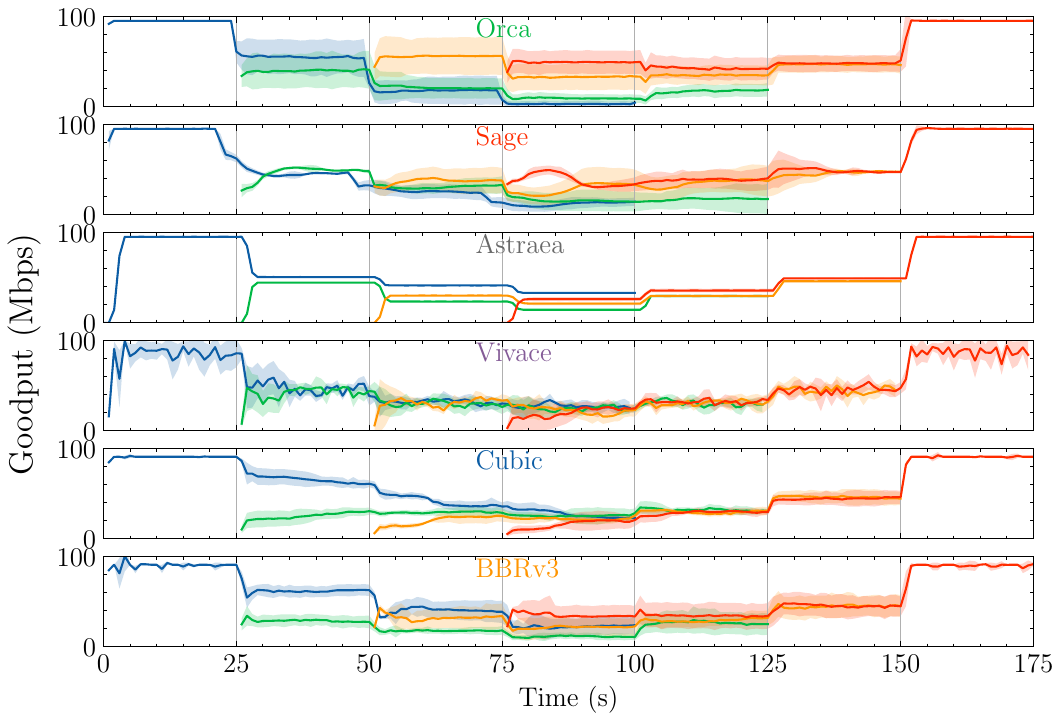}
        \caption{RTT: 200ms, Buffer Size: $4\times$ BDP}
        \label{fig:conv_200_4}
    \end{subfigure}
    \caption{\textbf{Convergence}. Goodput evolution for four competing flows in a dumbbell topology. Bottleneck capacity is $100$Mbps, flows experience the same base RTT of $200$ms, buffer capacity is set to $0.2\times$, $1\times$ and $4\times$ the BDP.}
    \label{fig:conv_200}
\end{figure*}

\noindent\textbf{Varying Bandwidth and Non-Congestive Loss}. We repeat the experiment above, but now update the bottleneck's random loss probability and bandwidth by uniformly selecting a value from the range $0$\% - $5$\% and $1$Mbps - $100$Mbps, respectively (base RTT is set to 100ms). In Figure \ref{fig:response_goodput_cdf}, the results of this experiment are shown in dashed lines. Orca, Sage and Astraea are resistant to random loss. This, in combination with the fact that we do not change the underlying base RTT in this experiment, means that they all perform slightly better compared to the non-loss setup discussed above. Once again, in \cite{pccvivace}, Vivace is shown to perform well with respect to non-congestive loss. However, in our study, we experiment with a much higher loss rate ($0$ to $5$\%) than in its evaluation in \cite{pccvivace} ($0$ to $1$\%), and this is why Vivace performs worse here (still, better than all the RL-based CC schemes). Cubic is known to perform very poorly in the presence of non-congestive loss, being a loss-based protocol; this is clear in Figures \ref{fig:response_goodput_cdf} and \ref{fig:response_send_rate}  where Cubic's sending rate is suppressed due to the recurring non-congestive loss. BBRv3's sending rate collapses in the presence of substantial non-congestive loss (Figure \ref{fig:response_send_rate}). This is because the average loss rate in this experiment exceeds the 2\% loss threshold that BBRv3 uses as a higher bound of bytes in flight \cite{bbrv3slides}.

\subsection{Convergence}
\label{convergence}

Figure \ref{fig:conv} illustrates convergence for the selected CC approaches; the experimental setup is the one used to measure efficiency (Section \ref{efficiency}). Orca converges to (relatively) stable goodput values quickly, for all buffer capacity values, but it is obvious that the bandwidth allocation is unfair, as also pointed out in Section \ref{fairness}. Similarly, Sage's convergence is swift even when the buffer capacity is high (Figure \ref{fig:conv_4}), but the stable bandwidth allocation is not always fair, with disturbances occurring when new flows join the network. Astraea’s convergence is remarkably swift and stable, having little to no variance among the different runs for each different setup. However, not all flows converge to a fair allocation. This contradicts the results shown in \cite{astraea}, where fast convergence and extremely fair bandwidth allocation was prevalent. Although we cannot be certain why this is the case, we believe that this was the result of applying fair queueing in the bottleneck as indicated in the configuration files made publicly available by the authors. Vivace shows a noisy and unstable convergence profile (albeit more fair than Orca and Sage) for all tested buffer capacity values (Figure \ref{fig:conv}); similar findings were shown in \cite{astraea}. Cubic flows converge to a fair share when the buffer capacity is $0.2\times$ and $1\times$ the BDP within the $25$-second interval, before a new flow enters the network. When the buffer capacity is $4\times$ the BDP, more time is needed for each flow to converge, as seen in Figure \ref{fig:conv_4}. BBRv3 convergence is greatly affected by buffer capacity, as discussed in Section \ref{fairness_rtt}; as the capacity increases, convergence gets slower and less fair.

In Figure \ref{fig:conv_200}, we plot the results of the same experiment but with the base RTT values set to $200$ms. This value is outside the training parameters of Astraea and Sage. Orca, as with the $20$ms case, is very unfair, with some of the competing flows getting close to zero bandwidth. Sage performs better despite the fact that it operates outside its training range; interestingly, one can see that goodput patterns are quite similar to Cubic (for all buffer capacities). This is to be expected because Cubic is one of the schemes used to train Sage. Astraea is, again, quick to converge to stable bandwidth allocations, but these are very unfair, with the flow started first being very aggressive to all subsequently joining flows. The performance of Cubic and BBRv3 is as expected, with even longer convergence times compared to the $20$ms case. Figures \ref{fig:conv_4} and \ref{fig:conv_200_4} show very similar patterns because the BDP values in the respective experiments are very close to each other.

%% file: sections/Related_Work.tex
\section{Related Work}
\label{related_work}

Table \ref{table:rltable} shows RL-based CC approaches proposed to date.

\noindent\textbf{Hybrid vs clean-slate approaches}. 
Hybrid approaches, those that incorporate existing heuristic CC algorithms during training or operation, have attracted increasing attention. The authors of \cite{sage} make the point that although existing schemes are not effective in all scenarios, they nonetheless embody valuable heuristics that can inform learning-based approaches. In the same spirit, Mutant \cite{mutant} extends this idea by using an online RL agent that continuously switches between existing CC algorithms, selecting the best performer for the observed network context. It formulates selection as a contextual N-armed bandit on a top-k pool chosen offline, then performs state-preserving ``mutations'' (the state of the current CC schemes is frozen and a new, better performing scheme is selected) when the estimated bandit reward favours another scheme. Another recent example of an RL-based congestion control scheme that leverages heuristic control is ORC \cite{orc}. ORC is an online reinforcement learning scheme that operates in conjunction with BBR to prevent the RL agent from making online exploratory decisions that could regress performance relative to the baseline BBR controller.

Clean-slate approaches have also made advances, a key contribution being Jury \cite{jury}. The aim of Jury is to achieve fairness across a much wider parameter range, motivated by the limitations of Astraea. Jury deliberately removes bandwidth/throughput features from the RL state. Its model is much simpler than that of previous works, with the only inputs to the RL model being changes in normalised RTT and loss. The model outputs a decision range rather than a concrete rate change. In a subsequent post-processing phase, a rate adjustment is made within the decision range given by the model, using the bandwidth occupancy ratio, a metric that can infer how much of the bottleneck a flow is occupying \cite{jury}. Evaluation shows that it is effective far beyond the range of training parameters seen during training, compared to the state-of-the-art Astraea.

\begin{table}[t]
\centering
\begin{tabularx}{\linewidth}{@{}lXX@{}}
\toprule
\textbf{Approach} & \textbf{Implementation} & \textbf{Open Sourced} \\ 
\midrule
Aurora \cite{jay2019deep} & UDT & Yes \\
DRL-CC \cite{xu2019experience}, Jury \cite{jury} & Linux Kernel & No, No \\
Orca \cite{abbasloo2020classic}, Astraea \cite{astraea}, Sage \cite{sage} & Linux Kernel & Yes, Yes, Yes \\
Spine \cite{tian2022spine}, Mutant \cite{mutant} & Linux Kernel & No, Yes \\
Eagle \cite{emara2020eagle}, Pareto \cite{emara2022pareto} & C++/Python & No, No \\
MOCC \cite{ma2022multi}, DeepCC \cite{deepcc} & UDT+CCP & No, Yes\\
ORC \cite{orc} & Pantheon & No \\
\bottomrule
\end{tabularx}
\setlength{\belowcaptionskip}{-10pt}
\caption{RL-based CC schemes implemented in emulation environments}
\label{table:rltable}
\end{table}

%% file: sections/Discussion.tex
\section{Discussion}
\label{conclusions}
In this paper, we presented a large-scale in-depth study of the latest RL-based CC approaches. To support verifiability and encourage transparent experimentation, we have made both our experimental codebase and the extracted datasets publicly available. In the following, we highlight the key findings, discuss the importance of openness in RL-driven CC research, and hope that our insights help guide the community toward the right directions.

\noindent\textbf{Embedding fairness into the reward function}. Prior to Astraea, fairness was a key shortcoming of RL-based congestion control schemes, as no existing method explicitly incorporated fairness into the reward formulation. For example, hybrid schemes such as Orca do not optimise fairness directly; rather, they inherit the AIMD-like fairness properties of Cubic. Other RL-based approaches either overlook fairness altogether or fail to encode it as part of the learning objective. Astraea addresses this by explicitly including fairness in the reward function, together with a multi-agent actor-critic training framework. As demonstrated in Section \ref{fairness}, Astraea achieves consistently high fairness in intra- and inter-RTT scenarios compared to other schemes.

\noindent\textbf{Unresponsiveness to dynamic networks}. The literature surrounding RL schemes explores responsiveness in terms of bandwidth, with experimentation often including emulations or in-the-wild cellular experiments \cite{astraea, abbasloo2020classic, jury}. Our results show that existing RL-based CC schemes struggle when the base RTT changes. The reward formulations of Astraea, Orca and Sage rely on minimum RTT or maximum bandwidth estimates, which can become outdated under changing conditions and mislead the policy. In addition, the traces used in training Sage, and the online environments used to train Astraea and Orca included only fixed-step changes in bandwidth but no changes in RTT, which limit its ability to perform well in such scenarios, typical of LEO satellite networks. Achieving responsiveness therefore requires a careful balance between training environments and a reward formulation that does not rely solely on the minimum observed RTT.

\noindent\textbf{Generalisation of performance to unseen network conditions}. Our study shows that RL-based congestion control schemes still struggle to generalise performance characteristics to unseen network conditions, for instance (i) Orca struggles to explore bandwidth values even slightly beyond its training range, (ii) Astraea demonstrates strong generalisation in terms of bandwidth and RTT, operating effectively far outside its training parameters, but its fairness properties do not carry over in such scenarios, (iii) Sage manages to function across a wider space, but its behaviour remains noisy and unstable. These findings highlight that, although RL schemes are effective within their domains, their generalisation remains narrow and highly dependent on the design of training regimes and reward functions.